\documentclass[10pt,journal]{IEEEtran}
\usepackage{graphicx}
\usepackage{amsmath}
\usepackage{amssymb}

\usepackage{caption}
\usepackage{subcaption}
\usepackage{color}
\usepackage{float}

\newtheorem{remark}{Remark}
\newtheorem{definition}{\bfseries{Definition}}
\newtheorem{lemma}{\bfseries{Lemma}}
\newtheorem{corollary}{\bfseries{Corollary}}

\newtheorem{problem}{\bfseries{Problem}}
\newtheorem{theorem}{\bfseries{Theorem}}

\makeatletter
\newcommand*\titleheader[1]{\gdef\@titleheader{#1}}
\AtBeginDocument{%
	\let\st@red@title\@title
	\def\@title{%
		\bgroup\normalfont\large\centering\@titleheader\par\egroup
		\vskip1.5em\st@red@title}
}
\makeatother

\titleheader{USC Technical Report}
\title{Robustness of Uncertain Switching Nonlinear Feedback Systems against Large Time-Variation}
\author{Yu-chen Sung, Sagar V. Patil, and Michael G. Safonov\thanks{Yu-chen Sung, Sagar V. Patil, and Michael G. Safonov are with the Department of Electrical Engineering - Systems, University of Southern California, Los Angeles, CA 90089-2563, USA. E-mail: yuchens@usc.edu, sagarvpa@usc.edu, and msafonov@usc.edu  \newline Manuscript submitted October 17, 2019 to {\em IEEE Trans. Automatic Control.}}}
\begin{document}
\maketitle
\thispagestyle{empty}
\pagestyle{empty}

%------------------------------------------------- Abstract -------------------------------------------------------------------------------------
\begin{abstract}
	For a non-linear MIMO feedback system, the robustness against uncertain time-variations in the feedback loop is investigated in an input-output framework. A general sufficient condition in terms of a bound on average rates of time-variation for the system to be stable is derived. The condition gives a tolerable limit on infrequent large variations or slow time-variation rate of a non-linear MIMO adaptive switching system.
\end{abstract}

%------------------------------------------------- Introduction --------------------------------------------------------------------------------
\section{Introduction}
%=====================================================
\par Typically, a feedback system having a time-varying non-linear loop function is equivalent to a feedback system with its loop function that switches among a family of time-invariant functions. These time-invariant functions are referred as frozen-time loop functions where each of them represents the loop function frozen at one certain time instant during the switching sequence. If the distances between all the frozen-time loop functions and the nominal loop function model, if exists, are bounded by a constant, then one can use classical small-gain theorem \cite{zames1966input} to determine the stability of the feedback system. However, in a more general case where the loop function is persistently time-varying and such a nominal loop function model does not exist, small-gain theorem does not conclude the stability of the feedback system. On the other hand, sufficient conditions for preserving stability of time-varying feedback systems in such general cases have been derived in the past in terms of maximum differences between consecutive frozen-time loop functions with various assumptions. Desoer \cite{desoer1970slowly} considered the discrete-time case where frozen-time functions are Hurwitz matrices which are linear and memoryless. In \cite{Solo_94}, Solo relaxed one of Desoer's assumptions for continuous-time cases where frozen-time functions are matrices which are not all Hurwitz. Zames and Wang \cite{Zames_Wang_91} improved Desoer's result by considering the slightly more general linear case in which frozen-time loop functions are assumed to be bounded, exponentially stabilizing, and time-invariant convolution operators.
\par The results in \cite{desoer1970slowly,Solo_94}, and \cite{Zames_Wang_91} are relevant for stability analysis of adaptive control when plants have large time-variations. An adaptive control logic tries to preserve stability and meet additional performance specifications simultaneously. In this context, a multiple model adaptive control \cite{anderson2000multiple,hespanha2001multiple} is developed to return a model close to the perturbed plant and corresponding designed controller. A system with Hysteresis Logic (HL) that switches controllers based on their real-time data-driven performance evaluations is developed by Morse, Mayne, and Goodwin in \cite{morse1992applications}, while a model-based switched system with HL that takes into account plant uncertainties or noise is investigated in \cite{narendra1997adaptive,baldi2010multi}, and \cite{angeli2002lyapunov}. A HL based switched system with an additional reset feature which safely discards past evaluated controller performances is developed by Battistelli, Hesphana, Mosca, and Tesi in \cite{BHMT}. The results of \cite{BHMT} are generalized in \cite{patil2016nonlinear} by (i) considering non-linear plants and controllers and (ii) adding a bumpless switching feature. A switching system that considers real-time and data-driven controller performance based on loop-shape specifications is developed in \cite{sung2018data}.
\par In a feedback system having an adaptive switching controller and a time-varying plant to be compensated, switching among controllers essentially leads to switching among loop functions. Therefore, the results in \cite{desoer1970slowly,Solo_94}, and \cite{Zames_Wang_91} can be used to analyze stability of adaptive switching systems. However, to investigate the impact of plant variations on stability of an adaptive switching control system, it is impractical to assume that either the adaptive loop function is linear or all the frozen-time loop functions are stabilizing. Adaptive loop functions are inherently non-linear. For example, in the HL switching algorithm \cite{morse1992applications}, the non-linear maximum operator causes adaptive loop functions to be non-linear. Also, switching algorithms may momentarily insert a destabilizing controller in the loop causing the resultant frozen-time loop function to be destabilizing \cite{dehghani2007unfalsified}. Moreover, the results in \cite{desoer1970slowly} and \cite{Zames_Wang_91} are compatible to slowly time-varying loop functions as they consider maximum difference between consecutive frozen-time loop functions, but they are not compatible to the adaptive loop functions having infrequent and large time-variations. \iffalse The same assumption that loop functions do not vary abruptly also limits certain classical stability criteria from being applied to determining stability of non-linear feedback systems. For example, the circle criterion \cite{haddad2011nonlinear} returns a sufficient condition in terms of ranges of non-linear feedback gain for asymptotic stability, and thus is not able to conclude stability if the gain is not within the range momentarily.\fi The recent works \cite{Shamma2012} and \cite{Sarwar08} too investigated stability of feedback systems with time-varying loop functions considering at least one of the assumptions on frozen-time loop functions that they are (i) stabilizing all the time, (ii) linear, and (iii) slightly different from adjacent frozen-time loop functions.
\par Therefore, we aim to solve a problem of determining under what condition in terms of average time-variation rate of time-varying non-linear loop function, stability of non-linear feedback system can be preserved without the three aforementioned assumptions on frozen-time loop functions.

\par The organization of the present paper is as follows. The preliminary facts are given in Section II. The problem formulation is described in Section III, followed by the main results in Section IV. \iffalse Section V discusses a limitation of a circle criterion.\fi A comparison of the results in the present paper with those in \cite{Zames_Wang_91} is discussed in Section V, followed by an application for adaptive control in Section VI. Two simulation examples are presented in Section VII, followed by conclusions in Section VIII.

\section{Preliminaries}
\par In the present paper, we consider discrete-time signals and systems. The sets of integers and real numbers are denoted by $\mathbb{Z}$ and $\mathbb{R}$ respectively. The transpose and the Euclidean norm are denoted by $\left[\,\cdot\,\right]'$ and $|\cdot|$ respectively. We denote the spectral radius of a matrix $A$ as $\lambda_{\text{max}}(A)$.
%=====================================================
%\par In the section, a $\sigma-$weighted norm for difference between time-varying nonlinear systems and their frozen-time nonlinear models proposed in \cite{Zames_Wang_91} is reviewed. The importance of the difference is reviewed in open-loop and closed-loop cases. First consider the following definitions:
%We develop an upper bound on the norm when average of $N$ samples of rate of variation of systems is considered. The upper bound generalized Proposition 2.2 of \cite{Zames_Wang_91} by considering average form of rate of variation of nonlinear systems.
%===========================
%\begin{definition}[Current Time]
%$\tau \in \mathbb{Z}$ is defined as the current time unless $\tau$ is defined otherwise, where $\mathbb{Z}$ is the space of integers. $\hfill \IEEEQEDopen$
%\end{definition}
%===========================
\begin{definition} \label{signal} \normalfont (\textit{Signal}): A real-valued function $x(t)$ of time $t$ is said to be a \emph{signal} mapping $t \in \mathbb{Z}$ to $x \in \mathbb{R}^n$, where $n \in \mathbb{Z}_{+}\backslash \{0\}$. \hfill $\square$
\end{definition}
%===========================
%\begin{definition}
%\label{CTsigmapsignalnorm} \normalfont (\textit{continuous-time \textit{$l_{\sigma p e}$ norm}}): Given $\sigma \geq 1$ and $p \in \left[1,\infty \right]$, for any continuous-time signal $x$ and for all $t_{1}, t_{2} \in \mathbb{R}$ where $t_{1}<t_{2}$, the \emph{moving-window fading-memory $l_{\sigma p}-$semi norm} is defined as
%\begin{align*}
%& \left\|x\right\|_{\sigma p,[t_1, t_2]} \triangleq
%	\begin{cases}
%	\sup\limits_{t \in [t_1,t_2]}
%	\sigma^{-(t_2-t)}\left|x(t)\right|, &\text{ if }p=\infty,
%----------
%\\
%	\left[\int_{t=t_1}^{t_2} \sigma^{-p(t_2-t)}\left|x(t)\right|^p  d(t) \right]^{\frac{1}{p}}, &\text{ otherwise,}
%	\end{cases} 
%	\end{align*}
%	We denote the $l_{\sigma p}$-semi norm as (i) $\left\|x\right\|_{\sigma p,t}$ if $t_1=-\infty$ and $t_2=t$ and (ii) $\left\|x\right\|_{p,[t_1,t_2]}$ if $\sigma=1$. The \emph{extended space} \emph{$\mathcal{L}_{\sigma p e}$} is defined as $\mathcal{L}_{\sigma p e} \triangleq \left\{x:\,\left\|x\right\|_{\sigma p,t}<\infty, \, \forall t \in \mathbb{R} \right\}$
%	$\hfill \square$
%\end{definition}
%===========================
\begin{definition}
	\label{sigmapsignalnorm} \normalfont (\textit{\textit{Signal Norm}}): Given $\sigma \geq 1$ and $p \in \left[1,\infty \right]$, for any signal $x$ and for all $t_{1}, t_{2} \in \mathbb{Z}$, where $t_{1}<t_{2}$, the \emph{moving-window fading-memory ${\ell}_{\sigma pe}-$semi norm} \cite{BHMT,Desoer_book,Sandburg_64} is defined as
	\begin{align*}
	& \left\|x\right\|_{\sigma p,[t_1, t_2]} \triangleq
	\begin{cases}
	\left[\sum_{\tau=t_1}^{t_2} \sigma^{-p(t_2-\tau)}\left|x(\tau)\right|^p \right]^{\frac{1}{p}}\hspace{-0.14cm}, \hspace{-0.285cm} &\text{ if }p\in\left[1,\infty\right) \hspace{-0.05cm},
	%----------
	\\
	\sup\limits_{\tau \in [t_1,t_2]}
	\sigma^{-(t_2-\tau)}\left|x(\tau)\right|,\hspace{-0.285cm} &\text{ if }p=\infty,
	\end{cases} 
	\end{align*}
	where $\left|\cdot\right|$ denotes the Euclidean norm. For brevity, the notation $\left\|x\right\|_{\sigma p,[t_1, t_2]}$ is simplified as (i) $\left\|x\right\|_{\sigma p,t}$ if $t_1=-\infty$ and $t_2=t$ and (ii) $\left\|x\right\|_{p,[t_1,t_2]}$ if $\sigma=1$. The \emph{extended space} \emph{$\ell^n_{\sigma p e}$} is defined as $\ell^n_{\sigma p e} \triangleq \left\{x:\,x(t) \in \mathbb{R}^n, n \in \mathbb{Z}_+\backslash \{0\} \text{ and }\left\|x\right\|_{\sigma p,t}<\infty, \, \forall t \in \mathbb{Z} \right\}$. $\hfill \square$
\end{definition}
%===========================
\begin{lemma} \label{pnormInq}
	For any signal $x$, it holds $\forall p \in [1,\infty], \forall \sigma \geq 1, \forall \tau \in \mathbb{Z},$ and $\forall t \leq \tau$ that $\left\|x\right\|_{\sigma p, t} \leq \sigma^{\tau-t} \left\|x\right\|_{\sigma p, \tau}$.
\end{lemma}	
\emph{Proof: }
Refer \cite{BHMT} and \cite{patil2016nonlinear}.
$\hfill \blacksquare$
%===========================
\begin{definition}
	\label{system} \normalfont (\textit{System}): Given $\sigma \geq 1$ and $p \in [1,\infty]$, a \textit{system} or \textit{operator} $H$ with input $u$ and output $y$ is a mapping $u^n \in \ell_{\sigma pe}$ to $y \in \ell^m_{\sigma pe}$, where $n,m\in \mathbb{Z}_+\backslash \{0\}$. $\hfill \square$
\end{definition}
%===========================
\begin{definition}
	\label{sigmapsysnorm} \normalfont (\textit{System Norm}): Given $\sigma \geq 1$ and $p \in \left[1,\infty \right]$, the \emph{moving-window fading-memory $\ell_{\sigma pe}$-semi norm} of a system $H$ with input $u$ is defined as $\|H\|_{\sigma p} \triangleq \sup_{\tau \in \mathbb{Z}}\left\|H\right\|_{\sigma p,\tau}$ if the supremum exists, else $\|H\|_{\sigma p} \triangleq \infty$,
	where
	\begin{align*}
	\left\|H\right\|_{\sigma p,\tau} \triangleq\sup_{\left\|u\right\|_{\sigma p,\tau}> 0} \frac{\left\|Hu\right\|_{\sigma p,\tau}}{\left\|u\right\|_{\sigma p, \tau}}
	\end{align*}
	is the norm of system $H$ at time $\tau \in \mathbb{Z}$. For simplicity, $\left\|H\right\|_{p} \triangleq \left\|H\right\|_{\sigma p}$ when $\sigma = 1$. \hfill $\square$
\end{definition}
%===========================
\begin{definition}
	\label{lsigmapestablesys} \normalfont (\textit{Stability and Degree of Stability}): Given $\sigma \geq 1$, $p \in [1, \infty]$, a system $H$ is said to be \textit{weakly $\ell_{\sigma pe}$-stable} if there exist a constant $c \in \mathbb{R}_+$ and an infinite time sequence $\left\{t_{i}: t_{i-1} < t_{i}, i \in \mathbb{Z}\right\}$ with $t_i \rightarrow \infty$ as $i \rightarrow \infty$ such that
	\begin{align}
	\|H\|_{\sigma p, t_i} \leq c, \quad \forall i \in \mathbb{Z} \label{sigmape_stability}
	\end{align}
	If $\{t_{i}\} = \mathbb{Z}$ then $H$ is said to be \textit{$\ell_{\sigma pe}$-stable} which we denote as $\ell_{\infty e}$-stability for $\sigma = 1$ and $p=\infty$. Given a system $H$, the supremum of the set of $\sigma$ for which $(\ref{sigmape_stability})$ holds is called the \textit{degree} $\sigma_{0}$ of stability of $H$. $\hfill \square$
\end{definition}
%===========================
\begin{remark} \label{poleradis}
	By \cite{BHMT} and \cite{patil2016nonlinear}, if a linear and time-invariant system $H$ has finite $\ell_{\sigma 2 e}$-semi norm with degree $\sigma\geq 1$, then $H$ has all its poles within the circle of radius $\frac{1}{\sigma}$. $\hfill \square$
\end{remark}
%===========================
\begin{remark} \label{morestable}
	If $\sigma_1 > \sigma_2$, where $\sigma_1 \triangleq \arg\sup_{\sigma\geq 1}\left\|H_1\right\|_{\sigma p}<\infty$ and $\sigma_2 \triangleq \arg\sup_{\sigma\geq 1}\left\|H_2\right\|_{\sigma p}<\infty$, then system $H_1$ is comparatively more stable than system $H_2$ by \cite{patil2016nonlinear}. $\hfill \square$
\end{remark}
%===========================
%\begin{definition}
%\label{CTlsigmapestablesys} \normalfont (\textit{stability of continuous-time systems}): Given $\sigma \geq 1$, $p \in [1, \infty]$, and time sequence $\left\{t_{i}: t_{i-1} < t_{i}, i \in \mathbb{Z}\right\}$, a continuous-time system $H$ is said to be \textit{weakly $l_{\sigma pe}$-stable} at $\{t_{i}\}$ if for all $u \in \mathcal{L}_{\sigma p e}$ we have
%\begin{align*}
%\|Hu\|_{\sigma p, t_i} \leq c \|u\|_{\sigma p, t_i} + \alpha, \quad \forall i \in \mathbb{Z}
%\end{align*}
%where $c\in\mathbb{R}_{+}$ and $\alpha \in\mathbb{R}_{+}$. If $\{t_{i}\} = \mathbb{R}$ then $H$ is said to be \textit{$l_{\sigma pe}$-stable} which we denote as $l_{2 e}$-stability for $\sigma = 1$ and $p=2$. $\hfill \square$
%\end{definition}
\begin{remark} 
	The adaptive switching control with reset mechanism, proposed in \cite{BHMT} and \cite{patil2016nonlinear} adaptively generates an infinite time sequence $\left\{t_{k}: t_{k-1} < t_{k}, k \in \mathbb{Z}\right\}$ with $t_k \rightarrow \infty$ as $k \rightarrow \infty$. By assuming finite-order plant and linear time-invariant controllers, it is proved in \cite[Theorem 1]{BHMT} that the adaptive switching control \cite{BHMT} preserves $\ell_{\infty e}$-stability. On the other hand, by relaxing these assumptions, it is proved in \cite[Theorem 3]{patil2016nonlinear} that the adaptive switching control \cite{patil2016nonlinear} too preserves $\ell_{\infty e}$-stability. $\hfill \square$
\end{remark}
%===========================
\begin{definition}
	\label{truncation and delayed} \normalfont (\textit{Backward Shift and Truncation Operators}): The operator $\mathcal{T}$ is defined as \emph{the backward shift operator} by
	\begin{align*}
	(\mathcal{T}^{\theta}x)(t)=x(t-\theta)
	\end{align*}
	for all $x \in \ell^n_{\sigma p e}$, $t \in \mathbb{Z}$, $n \in \mathbb{Z}_+\backslash \{0\}$ and $\theta \in \mathbb{Z}$. The operator $P_{\tau}$ is defined as \emph{the truncation operator} by
	\begin{align*}
	\left(P_{\tau}x\right)(t)=\begin{cases}
	x(t), \forall t \leq \tau,
	\\
	0, \text{ otherwise.}
	\end{cases}
	\end{align*}
	for all $\tau \in \mathbb{Z}$. $\hfill \square$
\end{definition}
%===========================
\begin{definition}
	\label{TI_Causal_Memoryless} \normalfont (\textit{Time-Invariant, Causal, and Memory-Less Systems}):
	A system $H$ is said to be (i) \emph{time-invariant} (TI) if $H\mathcal{T}=\mathcal{T}H$, (ii) \textit{causal} if $P_{t}H=P_{t} H P_{t}$, $\forall t \in \mathbb{Z}$, and (iii) \textit{memory-less} if $\left(P_{t}-P_{t-1}\right)H=\left(P_{t}-P_{t-1}\right) H \left(P_{t}-P_{t-1}\right),\forall t \in \mathbb{Z}$.  $\hfill \square$
\end{definition}
%===========================
\begin{definition} \label{snapshot} \normalfont (\textit{Frozen-Time Snapshots and Frozen-Time Extensions of Systems}):
	Consider a non-linear system $H$ with input $u \in \ell^n_{\sigma pe}$ and output $Hu \in \ell^m_{\sigma pe}$ where $n,m \in \mathbb{Z}_+\backslash \{0\}$. The \emph{frozen-time snapshot} $h_{\tau}:\mathbb{\ell}^n_{\sigma p e} \mapsto \mathbb{R}^m$ of $H$ at time $\tau \in \mathbb{Z}$ is defined by $h_{\tau} u = \left(Hu\right)(\tau)$. The unique \emph{frozen-time extension} $H_{\tau}:\mathbb{\ell}^n_{\sigma p e} \mapsto \mathbb{\ell}^m_{\sigma p e}$ of $H$ at time $\tau \in \mathbb{Z}$ is defined by $\left(H_{\tau}u\right)(t) = h_{\tau} \mathcal{T}^{\tau-t}u$ for all $t \in \mathbb{Z} $. The difference between $h_{t-1}$ and $h_{t}$ is denoted as $\nabla h_{t} \triangleq h_{t-1} - h_{t}$, and the difference between $H$ and $H_{t}$ is denoted as $\nabla H_{t} \triangleq H-H_{t}$ for all $t\leq \tau$.
	 
	\hfill $\square$
\end{definition}
%===========================
\begin{remark} 
	In Definition \ref{snapshot}, the frozen-time extension $H_{t}$ at $t\in \mathbb{Z}$ is TI.	$\hfill \square$
\end{remark}
%===========================
\begin{lemma} \label{snapshot_norm_bound}
	Given a non-linear system $H$ with input $u\in \ell^n_{\sigma \infty e}$, where $n \in \mathbb{Z}_+\backslash \{0\}$, we have
	\begin{align*}
	\max_{t \in \mathbb{Z}}\left\|h_t\right\|_{\sigma \infty} = \left\|H\right\|_{\sigma \infty}.
	\end{align*}
\end{lemma}
\emph{Proof: } By Definitions \ref{sigmapsysnorm} and \ref{snapshot}, we get 	
\begin{align*}
\left\|H\right\|_{\sigma \infty} \triangleq 
& \max_{\substack{t \in \mathbb{Z} \\ \quad \,\, u\in \ell^n_{\sigma \infty e}} }\frac{\left\|Hu\right\|_{\sigma \infty,t}}{\left\|u\right\|_{\sigma \infty,t}} 
=
\max_{\substack{t \in \mathbb{Z} \\ \quad \,\, u\in \ell^n_{\sigma \infty e}}}\frac{\left|\left(Hu\right)(t)\right|}{\left\|u\right\|_{\sigma \infty,t}}
\\
=
&\max_{t \in \mathbb{Z}}\left\|h_t\right\|_{\sigma \infty}.  \hspace{4.9cm} \blacksquare
\end{align*}
%===========================
\begin{figure}[h]
	\centering
	\vspace{0cm}
	\includegraphics[scale=1]{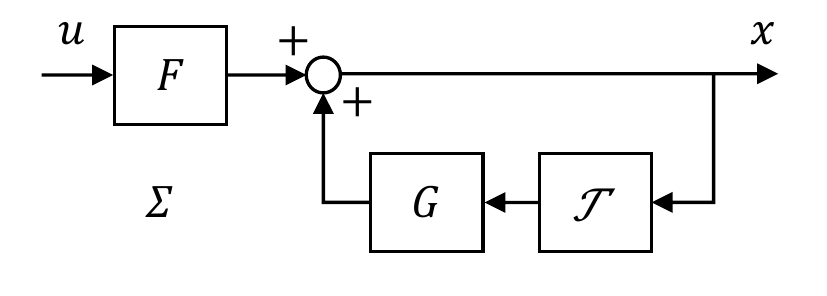}
	\caption{A general feedback system $\Sigma$.}
	\vspace{0cm}
	\label{feedbacksys}
\end{figure}
%===========================
\begin{lemma} \label{delAndela}
	Given a non-linear system $H$ with input $u$ and a pair of times $t, \tau \in \mathbb{Z}$, where $t \leq \tau$, then $\forall u \in \ell^n_{\sigma p e}$ with $n \in \mathbb{Z}_+\backslash \{0\}$, we have
	\begin{align*}
	\left(\nabla H_{\tau}u\right)(t)=\begin{cases} \left(\sum_{i=t+1}^{\tau}\nabla h_i\right)\mathcal{T}^{\tau-t}u, &\text{if }t < \tau, \\ 0, &\text{if }t=\tau. \end{cases}
	\end{align*}
\end{lemma}
\emph{Proof: }	
The lemma is an immediate consequence of Definition \ref{snapshot}.
$\hfill \blacksquare$
%===========================
\begin{remark} \label{dK}
	For a system $H$ with input $u \in \ell^n_{\sigma p e}$, where $n \in \mathbb{Z}_+\backslash \{0\}$, the reference \cite{Zames_Wang_91} defines the term $d_{\sigma}(H)$ by $d_{\sigma}(H) \triangleq \sup_{t \in \mathbb{Z},u \in {\ell}^n_{\sigma p e}}\left\|\mathcal{T}\left(Hu\right)(t)-\left(H\mathcal{T}u\right)(t)\right\|_{\sigma \infty}$, which is equal to $\sup_{\tau \in \mathbb{Z}} \sigma \left\| \nabla h_{\tau} \right\|_{\sigma \infty}$ according to our Definition \ref{snapshot}. $\hfill \square$
\end{remark}
%===========================
\par A system $H$ is said to be slowly time-varying when $\left\| \nabla h_{t} \right\|_{\sigma \infty}$ is small for all $t \in \mathbb{Z}$ and it is said to be infrequently varying over the interval $L$ when $\left\| \nabla h_{t} \right\|_{\sigma \infty}$ has small average over the interval $L$.
%===========================
\par The $N$-width average variation rate of a time-varying non-linear system is defined as follows.
%===========================
\begin{definition}
	\label{Nwidth} \normalfont (\textit{$N$-Width Average Variation Rate}):
	Given a causal non-linear system $H$ and an integer $N \in \mathbb{Z}_+\backslash \{0\}$, the \emph{$N$-width average variation rate} of $H$ is defined as
	\begin{align}
	d_{\sigma,N}(H)(t) \triangleq \frac{1}{N}\sum_{i=t-N+1}^{t} \left\|\nabla h_{i}\right\|_{\sigma \infty} \label{d_N}.
	\end{align}
	We define $\bar{d}_{\sigma,N}(H)$ as the \emph{least upper bound} on $d_{\sigma,N}(H)(t)$ for all $t \in \mathbb{Z}$, i.e.,
	\begin{align*}
	\bar{d}_{\sigma,N}(H) \triangleq \sup_{t \in \mathbb{Z}} d_{\sigma,N}(H)(t).
	\end{align*} $\hfill \square$
\end{definition} 
%===========================
\begin{remark}
	A special case of Definition \ref{Nwidth} having $N=1$ is discussed in \cite{Zames_Wang_91}. $\hfill \square$
\end{remark}
%=====================================================
\begin{lemma} \label{d_sigma_N_of_prod}
	Consider constants $\sigma>1$ and $N \in \mathbb{Z}_+ \backslash \{0\}$. Consider a system $H = GK$ where $K$ is a time-invariant non-linear system with finite $\left\|K\right\|_{\sigma \infty}$ and $G$ is a time-varying non-linear system. Let the frozen-time snapshots of $G$ and $H$ at time $t \in \mathbb{Z}$ be denoted by $g_t$ and $h_t$ respectively. Then we have
	\begin{align}
	\left\|\nabla h_t \right\|_{\sigma \infty} \leq \left\|\nabla g_t \right\|_{\sigma \infty} \left\|K\right\|_{\sigma \infty} \label{d_sigma_N_of_prod_inq1}
	\end{align}
	and
	\begin{align}
	\bar{d}_{\sigma,N}(H) \leq \left\|K\right\|_{\sigma \infty} \bar{d}_{\sigma,N}(G). \label{d_sigma_N_of_prod_inq2}
	\end{align}
\end{lemma}
%============================
\begin{figure}[ht]
	\centering
	\includegraphics[trim=0cm 0cm 0cm 0cm]{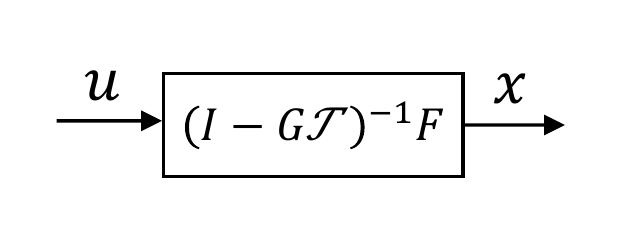}
	\vspace{-0.5cm}
	\caption{The feedback system $\Sigma=\left(I-G\mathcal{T}\right)^{-1}F$.}
	\vspace{-0cm}
	\label{I-GTinvF}
\end{figure}
%============================
\emph{Proof: }	
Let the input to the system $H$ be denoted by $x$. Then by Definitions \ref{sigmapsysnorm} and \ref{snapshot}, for all $t \in \mathbb{Z}$, we have
\begin{align*}
\left\|\nabla h_t x \right\|_{\sigma \infty,t}
= & \left\|h_{t-1}x - h_{t} x \right\|_{\sigma \infty,t}
\\
= & \left\|g_{t-1}\left(Kx\right) - g_{t} \left(Kx\right) \right\|_{\sigma \infty,t}
\\
\leq & \left\|\nabla g_t \right\|_{\sigma \infty} \left\|K\right\|_{\sigma \infty} \left\|x \right\|_{\sigma \infty,t}.
\end{align*}
Therefore, $(\ref{d_sigma_N_of_prod_inq1})$ holds by Definition \ref{sigmapsysnorm} and $(\ref{d_sigma_N_of_prod_inq2})$ is a consequence of $(\ref{d_sigma_N_of_prod_inq1})$ and Definition \ref{Nwidth}.  $\hfill \blacksquare$
%===========================
\iffalse
\begin{remark}
	A special case of Definition \ref{Nwidth} having $N=1$ is discussed in \cite{Zames_Wang_91}. $\hfill \square$
\end{remark}
\fi
%=====================================================
\section{Problem Formulation}
%===========================
\par We consider the general feedback system $\Sigma$ in Fig. \ref{feedbacksys}, which can be described as $\Sigma = \left(I - G\mathcal{T}\right)^{-1}F$ shown in Fig. \ref{I-GTinvF}, where $F$ and $G$ are causal non-linear operators. The main problem is formulated as follows.
%===========================
\begin{problem} \label{P1}
	Consider $\sigma\geq 1$. Consider the non-linear feedback system $\Sigma$ in Fig. \ref{I-GTinvF} where $G:\mathbb{\ell}^m_{\sigma \infty e} \mapsto \mathbb{\ell}^m_{\sigma \infty e}$ and $F:\mathbb{\ell}^n_{\sigma \infty e} \mapsto \mathbb{\ell}^m_{\sigma \infty e}$ with $n,m \in \mathbb{Z}_+\backslash \{0\}$ and  $\left\|F\right\|_{\infty} < \infty$. Given a time sequence $\left\{t_i:  t_{i-1} \leq t_i, i \in \mathbb{Z} \right\}$, find a sufficient condition such that, for all $i \in \mathbb{Z}$, the inequality $\|\Sigma\|_{\infty,t_{i}} \leq c$ holds for some constant $c > 0$. \hfill $\square$
\end{problem}
%===========================
\begin{remark}
	A solution to problem \ref{P1} implies the system $\Sigma$ is \emph{weakly ${\ell}_{\infty e}$-stable} with respect to a given time sequence $\left\{t_i\right\}$. In case $\left\{t_i\right\}=\mathbb{Z}$, the system $\Sigma$ is \emph{${\ell}_{\infty e}$-stable} as well. $\hfill \square$
\end{remark}
%===========================
\section{Main Results}
\par 
We derive a solution to the Problem $\ref{P1}$ without posing any assumptions on the time-varying MIMO feedback system in Fig. \ref{feedbacksys}. In the later part, we consider some special cases of the derived results as well as compare them with the results from Zames and Wang's paper \cite{Zames_Wang_91}.
%===========================
\begin{lemma} \label{ZamesLemma}
	Consider $\sigma, \sigma_{0} \in \mathbb{R}_{+}$ where $1 \leq \sigma < \sigma_{0}$. Consider two causal non-linear systems $H$ and $G$. Define
	\begin{align}
	&c_{\sigma,\sigma_0}(G,t)
	\triangleq 
	\sup_{i \geq 1}
	\left[\left(\frac{\sigma}{\sigma_0}\right)^i
	\sum_{q=t-i+1}^{t}\left\|\nabla g_{q}\right\|_{\sigma \infty} \right]. \label{c_sigma}
	\end{align}
	Then $\forall t \in \mathbb{Z}$, we have
	\begin{align}
	&\left\|h_t\nabla G_t\right\|_{\sigma \infty,t}
	\leq
	\left\|h_{t} \right\|_{\sigma_0 \infty,t} 
	c_{\sigma,\sigma_0}(G,t). \label{ZamesLemmaresult}
	\end{align}
	
\end{lemma}

\emph{Proof:}	Refer Appendix A. \hfill $\blacksquare$
%==============================================
%\begin{corollary} \label{ZamesLemmaCorollary}
%	Consider Lemma \ref{ZamesLemma}. Then 
%	\begin{align*}
%	\left\|h_t\nabla K_t\right\|_{\sigma \infty}
%	\leq & 
%	\left\|h_{t} \right\|_{\sigma_0 \infty} 
%	c_{\sigma,\sigma_0}(K,t).
%	\end{align*}
	
%	\emph{Proof :}\normalfont The corollary is a consequence of Definition \ref{sigmapsysnorm} and Lemma \ref{ZamesLemma}.\hfill $\blacksquare$
%\end{corollary}
%==============================================
\par The following lemma is derived for systems with bounded variation rate defined in Definition \ref{Nwidth}.
%==============================================
\begin{lemma} \label{c_Nlemma}
	Consider $\sigma, \sigma_{0} \in \mathbb{R}_{+}$ where $1 \leq \sigma < \sigma_{0}$. Consider a causal non-linear system $G$  having $\bar{d}_{\sigma,N}(G) \in \mathbb{R}_{+}$ for some $N \in \mathbb{Z}_+$. Define
	\begin{align}
	&c_{\sigma,N}(G)
	\triangleq 
	\left(e \ln\left(\frac{\sigma_0}{\sigma}\right)\right)^{-1}
	\left(\frac{\sigma_0}{\sigma}\right)^{N-1}
	\bar{d}_{\sigma,N}(G). \label{c_N}
	\end{align}
	Then for all $t \in \mathbb{Z}$, we have
	\begin{align}
	&c_{\sigma,\sigma_0}(G,t) \leq c_{\sigma,N}(G). \nonumber
	\end{align}
	
\end{lemma}

\emph{Proof:}	Refer Appendix B. \hfill $\blacksquare$
%==============================================
\begin{lemma} \label{ZamesLemmaNwidth}
	Consider $\sigma, \sigma_{0} \in \mathbb{R}_{+}$ where $1 \leq \sigma < \sigma_{0}$. Consider two causal non-linear systems $H$ and $G$ having $\bar{d}_{\sigma,N}(G) \in \mathbb{R}_{+}$ for some $N \in \mathbb{Z}_+\backslash \{0\}$. Then $\forall t \in \mathbb{Z}$, we have
	\begin{align}
	&\left\|h_t\nabla G_t\right\|_{\sigma \infty,t}
	\leq
	\left\|h_{t} \right\|_{\sigma_0 \infty,t} 
	c_{\sigma,N}(G). \nonumber
	\end{align}
	%and
	%\begin{align}
	%&\left\|h_t\nabla K_t\right\|_{\sigma \infty}
	%\leq
	%\left\|h_{t} \right\|_{\sigma_0 \infty} 
	%c_{\sigma,N}(K) \nonumber
	%\end{align}
	
	\emph{Proof :}\normalfont The lemma is a consequence of Lemma \ref{ZamesLemma} and Lemma \ref{c_Nlemma}.\hfill $\blacksquare$
\end{lemma}
%==============================================
\begin{remark}  In \cite{Zames_Wang_91}, it is derived that $\left\|h_t \nabla K_t\right\|_{\sigma \infty} \leq \left\|h_t\right\|_{\sigma_0 \infty} \sigma^{-1} \left(e \ln \left(\frac{\sigma_0}{\sigma}\right)\right)^{-1} d_{\sigma}(K)$. By Remark \ref{dK} and Definition \ref{snapshot}, $d_{\sigma}(K) = \sigma \sup_{\tau \in \mathbb{Z}} \left\|\nabla k_{\tau}\right\|_{\sigma \infty}$. Therefore Lemma \ref{ZamesLemmaNwidth} generalizes the result in \cite{Zames_Wang_91} by considering $N\geq 1$. $\hfill \square$
\end{remark}
%=============================================
\par A solution for Problem \ref{P1} is proposed as follows.

%\par Theorem \ref{xdelayed2x} gives a sufficient condition for system $\Sigma$ in Fig. \ref{I-GTinvF} to be weakly ${\ell}_{\infty e}$-stable at $\left\{t_i\right\}$ by considering frozen-time snapshot of $(I-G_tT)^{-1}$, which is the sensitivity function of the closed-loop system with the frozen-time snapshot extension $G_t$. A result similar to the sufficient condition $(\ref{Thm1C1})$ can be derived by considering complementary sensitivity $(I-G_tT)^{-1}G_tT$ as follows.
%=============================================
\begin{theorem} \label{stable_xdelayed2x_CS}
	Consider $\sigma, \sigma_{0} \in \mathbb{R}_{+}$ where $1\leq\sigma<\sigma_{0}$, $\rho \in \mathbb{R}_+$, and $\left\{t_{i} : t_{i-1} \leq t_{i}, i \in \mathbb{Z}\right\}$. Consider the non-linear feedback system $\Sigma$ in Fig. \ref{I-GTinvF} where $F:\mathbb{\ell}^n_{\sigma \infty e} \mapsto \mathbb{\ell}^m_{\sigma \infty e}$ having $\left\|F\right\|_{\infty} < \infty$ and $G:\mathbb{\ell}^m_{\sigma \infty e} \mapsto \mathbb{\ell}^m_{\sigma \infty e}$ with $n,m \in \mathbb{Z}_+\backslash \{0\}$. Let $s_t$ and $l_t$ be the TI frozen-time snapshots of $(I-G_{t}\mathcal{T})^{-1}$ and $(I-G_{t}\mathcal{T})^{-1}G_{t}\mathcal{T}$ respectively. Define
	\begin{align}
	\psi(t)&\triangleq \max\left\{
	\min\left\{
	\left\|l_t\right\|_{\sigma_0 \infty,t} c_{\sigma,\sigma_0}(G,t),
	\left\|g_t\right\|_{\sigma \infty,t}
	\right\}
	,\sigma^{-1} 
	\right\}, \label{psi}
	\\
	c &\triangleq \sigma^{\bar{t}-1}\frac{\beta}{1-\rho}, \label{c}
	\\
	\bar{t}&\triangleq \sup_{i \in \mathbb{Z}}\left(t_i-t_{i-1} \right), \label{bar_t}
	\\
	\beta & \triangleq \bar{t} \|F\|_{\infty} \sup_{i \in \mathbb{Z}} \left(\max_{t \in \left[t_{i-1}+1,t_i\right]} \left\|s_t\right\|_{\infty,t} \rho^{t_i-t}\right), \label{beta}
	\end{align}
	and $c_{\sigma,\sigma_0}(G,t)$ is defined in $(\ref{c_sigma})$.
	If
	\begin{align}
	\rho^{t_i-t} \geq \prod_{j=t+1}^{t_i} \psi(j),\quad \forall t \in\left[t_{i-1},t_i-1\right],\forall i \in \mathbb{Z}, \label{Thm2C1}
	\end{align}
	then for all $i \in \mathbb{Z}$ we have
	\begin{align}
	\|x\|_{\sigma \infty, t_i} \leq \rho^{t_i-t_{i-1}} \|x\|_{\sigma \infty, t_{i-1}} + \beta \|u\|_{\infty,t_i} \label{Thm2Result0}
	\end{align}
	Furthermore, if $\rho \in \left(\sigma^{-1},1\right)$, then for all $i \in \mathbb{Z}$ we have $\|\Sigma\|_{\infty,t_i} \leq c$.
\end{theorem}

\emph{Proof:} Refer Appendix C. \hfill $\blacksquare$
%=============================================
%\begin{remark} Due to $(\ref{psi})$ in Theorem \ref{stable_xdelayed2x_CS}, the sufficient condition $(\ref{Thm2C1})$ holds if both $\prod_{t \in (t_{i-1}+1,t_{i}]} \|l_t\|_{\sigma\infty,t}\,\forall i \in \mathbb{Z}$ and $\bar{d}_{\sigma,N}(G)$ are small enough. $\hfill \QED$
%\end{remark}
%=============================================
\begin{remark} 
	To hold condition $(\ref{Thm2C1})$, it is not necessary for the frozen-time snapshots $l_{t}$ to be stable for all time. It can be unstable at some times other than $\{t_{i}\}$ provided  $c_{\sigma,\sigma_0}(G,t)$ is small enough for all $t \in [t_{i-1}+1,t_i]$ and for all $i \in \mathbb{Z}$. \hfill $\square$
\end{remark}
%=============================================
\par Like Zames and Wang's sufficient condition \cite[inequality $(2.22)$]{Zames_Wang_91} for system $\Sigma$ to be $\ell_{\infty e}$- stable, Theorem \ref{stable_xdelayed2x_CS} considers the frozen-time snapshots $l_t$, but it does not consider assumptions that $F$ and $G$ are linear, $G$ is stabilizing, and $\left\{t_i\right\}=\mathbb{Z}$. Therefore, Theorem \ref{stable_xdelayed2x_CS} is a generalization of \cite[inequality $(2.22)$]{Zames_Wang_91}.
%=============================================
\begin{corollary} \label{stable_xdelayed2x_CS_c_N}
	Let $\left\{t_{i} : t_{i-1} \leq t_{i}, i \in \mathbb{Z}\right\}$ be a time sequence and let $\rho \in \left(0,1\right)$ be a constant. Define
	\begin{align}
	\psi_N(t)&\triangleq 
	\max\left\{ 
	\min\left\{
	\left\|l_t\right\|_{\sigma_0 \infty,t} c_{\sigma,N}(G),
	\left\|g_t\right\|_{\sigma \infty,t}
	\right\}
	,\sigma^{-1} 
	\right\}, \label{psi_N}
	\end{align}
	where $c$ is defined as in $(\ref{c})$ and $c_{\sigma,N}(G)$ is defined in $(\ref{c_N})$.
	If
	\begin{align}
	\rho^{t_i-t} \geq \prod_{j=t+1}^{t_i} \psi_N(j),\quad \forall t \in\left[t_{i-1},t_i-1\right],\forall i \in \mathbb{Z}, \label{Thm2_c_N_C1}
	\end{align}
	then
	\begin{align*}
	\|\Sigma\|_{\infty,t_i} \leq c, \quad \forall i \in \mathbb{Z}.
	\end{align*}
	%%%%%%%%%%
	
	\textit{Proof:} By Lemma \ref{c_Nlemma}, $c_{\sigma,\sigma_0}(G,t) \leq c_{\sigma,N}(G)$ for all $t \in \mathbb{Z}$, which implies $\psi(t)\leq \psi_N(t)$ for all $t \in \mathbb{Z}$. Therefore if $(\ref{Thm2_c_N_C1})$ holds then $(\ref{Thm2C1})$ holds. Hence, the corollary is proved. \hfill $\blacksquare$
\end{corollary}
%=============================================
\par The following corollary gives a sufficient condition for the system $\Sigma$ to be ${\ell}_{\infty e}$-stable for all time.
%================New theorem: $\ell_{\infty e}$-stability sufficient condition==========
\begin{corollary}\label{notweakstab}
	Let $\left\{t_{i} : t_{i-1} \leq t_{i}, i \in \mathbb{Z}\right\}$ be a time sequence and let $\rho \in \left(0,1\right)$ be a constant.
	Define
	\begin{align}
	\hat{c} &\triangleq \left(\frac{\left(\sigma \rho \right)^{\bar{t}}}{1-\rho}+1\right)\hat{\beta}, \label{globalc}
	\\
	%------------
	\hat{\psi}(t)&\triangleq
	\begin{cases}
	\max\left\{\sigma^{-1},\left\|l_t\right\|_{\sigma_0 \infty,t}c_{\sigma,\sigma_0}(G,t)\right\}, &\\ \qquad \text{if $G_t$ is stabilizing},
	\\
	\max\left\{\sigma^{-1},\left\|g_t\right\|_{\sigma \infty,t}\right\}, &\\ \qquad \text{if $G_t$ is destabilizing},
	\end{cases} \label{globalpsi}
	\\
	%------------
	\hat{\beta} & \triangleq 1+\left(\sigma\rho\right)^{\bar{t}}
	\frac{\rho}{ 1-\rho}
	\max_{t \in \mathbb{Z}}\gamma(t), \label{globalbeta}
	\\
	%------------
	\gamma(t) &\triangleq 
	\begin{cases}
	\left\|F\right\|_{\infty}\left\|s_t\right\|_{\sigma \infty,t}, &\text{if $G_t$ is stabilizing},
	\\
	\left\|F\right\|_{\infty}, &\text{if $G_t$ is destabilizing}
	\end{cases} \label{globalgamma}
	\end{align}
	and $\bar{t}$ is defined as in $\left(\ref{bar_t}\right)$. If
	\begin{align}
	\rho^{t_i-t} \geq \prod_{j=t+1}^{t_i} \hat{\psi}(j),\quad \forall t \in\left[t_{i-1},t_i-1\right],\forall i \in \mathbb{Z}, \label{Thm3C1}
	\end{align} 
	then \begin{align}
	\|\Sigma\|_{\infty,t} \leq \hat{c},\quad \forall t \in \mathbb{Z}. \label{stab}
	\end{align}
	
\end{corollary}
\emph{Proof:} Refer Appendix D. \hfill $\blacksquare$
%===========================
\begin{remark} \label{Thm3finitec}
	By $(\ref{globalgamma})$, $\gamma(t)$ is bounded for all $t \in \mathbb{Z}$ because $\left\|F\right\|_{\infty}<\infty$ and $\left\|s_t\right\|_{\sigma \infty,t}<\infty$ when $G_t$ is stabilizing. Therefore, by $(\ref{globalc})$, $(\ref{globalbeta})$, and $(\ref{globalgamma})$, $\hat{c}$ is bounded provided $\bar{t}$ is finite. $\hfill \square$
\end{remark}
%=============================================
%=============================================
\par The following lemma gives a sufficient condition for the system $\Sigma$ to be ${\ell}_{\infty e}$-stable for all time given the upper bound $\bar{d}_{\sigma, N}(G)$, on average variation rate of loop function $G$.
%================New theorem: $\ell_{\infty e}$-stability sufficient condition==========
\begin{lemma}\label{notweakstab_cN}
	Let  $\left\{t_{i} : t_{i-1} \leq t_{i}, i \in \mathbb{Z}\right\}$ be a time sequence and let $\rho \in \left(0,1\right)$ be a constant. Consider $\hat{c}$ defined in $(\ref{globalc})$. Define
	\begin{align}
	\hat{\psi}_N(t)&\triangleq
	\begin{cases}
	\max\left\{\sigma^{-1},\left\|l_t\right\|_{\sigma_0 \infty,t}c_{\sigma,N}(G)\right\}, &\\ \qquad \text{if $G_t$ is stabilizing},
	\\
	\max\left\{\sigma^{-1},\left\|g_t\right\|_{\sigma \infty,t}\right\}, &\\ \qquad \text{if $G_t$ is destabilizing}.
	\end{cases} \label{globalpsi_N}
	\end{align}
	If
	\begin{align}
	\rho^{t_i-t} \geq \prod_{j=t+1}^{t_i} \hat{\psi}_N(j),\quad \forall t \in\left[t_{i-1},t_i-1\right],\forall i \in \mathbb{Z}, \label{stab_cN}
	\end{align} 
	then \begin{align*}
	\|\Sigma\|_{\infty,t} \leq \hat{c},\quad\forall t \in \mathbb{Z}.
	\end{align*}
\end{lemma}
%%%%%%%%%%

\textit{Proof:} By Lemma \ref{c_Nlemma}, $c_{\sigma,\sigma_0}(G,t) \leq c_{\sigma,N}(G)$ for all $t \in \mathbb{Z}$, which implies $\hat{\psi}(t)\leq \hat{\psi}_N(t)$ for all $t \in \mathbb{Z}$. Therefore if $(\ref{stab_cN})$ holds then $(\ref{stab})$ holds. Hence, the lemma is proved. \hfill $\blacksquare$
%=============================================
\begin{lemma} \label{stable_xdelayed2x_CS_special}
	Consider $\sigma, \sigma_{0} \in \mathbb{R}_{+}$, where $1 \leq \sigma < \sigma_{0}$, and $\rho \in (\sigma^{-1},1)$. Consider the non-linear feedback system $\Sigma$ in Fig. \ref{I-GTinvF} where $F:\mathbb{\ell}^n_{\sigma \infty e} \mapsto \mathbb{\ell}^m_{\sigma \infty e}$ having $\left\|F\right\|_{\infty} < \infty$ and $G:\mathbb{\ell}^m_{\sigma \infty e} \mapsto \mathbb{\ell}^m_{\sigma \infty e}$ with $n,m \in \mathbb{Z}_+\backslash \{0\}$. Let $s_t$ and $l_t$ be the TI frozen-time snapshots of $(I-G_{t}\mathcal{T})^{-1}$ and $(I-G_{t}\mathcal{T})^{-1}G_{t}\mathcal{T}$ respectively such that $\sup_{t\in\mathbb{Z}}\|s_t\|_{\infty} < \infty$ and $\sup_{t\in\mathbb{Z}}\|l_t\|_{\sigma_0 \infty} < \infty$. Define
	\begin{align} 
	&\bar{c} \triangleq \frac{\|F\|_{\infty}\sup_{t \in \mathbb{Z}} \left\| {s}_t \right\|_{\infty}}{1-\rho} \label{bar_c}.
	\end{align} If
	\begin{align}
	c_{\sigma,\sigma_0}(G,t) \leq \frac{\rho}{\left\|l_t\right\|_{\sigma_0 \infty,t}} \quad \forall t \in \mathbb{Z} \label{CS_specical_c1}
	\end{align}
	then for all $t \in \mathbb{Z}$, we have $\|\Sigma\|_{\infty,t} \leq \bar{c}$.
\end{lemma}
\emph{Proof:} If $(\ref{CS_specical_c1})$ holds, then $(\ref{Thm2C1})$ holds too for the special case $t_i-t_{i-1}=1$ by $(\ref{c_sigma})$ and $(\ref{psi})$. If $\bar{t}=1$ in $(\ref{c})$, $(\ref{bar_t})$ and $(\ref{beta})$ then we get $\beta=\bar{t}\|F\|_{\infty}\sup_{t \in \mathbb{Z}} \left\| {s}_t \right\|_{\infty}$. Therefore, by $(\ref{bar_c})$ we have $c = \bar{c}$. By Theorem \ref{stable_xdelayed2x_CS}, $\|\Sigma\|_{\infty,t} \leq \bar{c}, \forall t \in \mathbb{Z}$. \hfill $\blacksquare$
%=============================================
\par The following corollary is derived from Lemma \ref{stable_xdelayed2x_CS_special} given the upper bound $\bar{d}_{\sigma, N}(G)$, on average variation rate of loop function $G$.
%=============================================
\begin{corollary} \label{stable_xdelayed2x_CS_special_Nwidth}
	Define \begin{align} 
	&\bar{\bar{d}}_{\sigma,N}(G) \triangleq
	\left(\frac{\sigma_0}{\sigma}\right)^{1-N}
	\left(e \ln \left(\frac{\sigma_0}{\sigma}\right)\right) 
	\left(\sup_{t \in \mathbb{Z}} \left\| l_t \right\|_{\sigma_0 \infty}\right)^{-1}
	\rho. \label{bar_bar_d}
	\end{align}
	Consider $\bar{d}_{\sigma,N}(G)$ defined in Definition \ref{Nwidth} and $\bar{c}$ defined in $(\ref{bar_c})$. If
	\begin{align}
	\bar{d}_{\sigma,N}(G) \leq \bar{\bar{d}}_{\sigma,N}(G), \label{CS_specical_c_N_c1}
	\end{align}
	then for all $t \in \mathbb{Z}$, we have $\left\|\Sigma\right\|_{\infty,t} \leq \bar{c}$.
\end{corollary}

\emph{Proof:} Let $(\ref{CS_specical_c_N_c1})$ holds, then
\begin{align*}
&\bar{d}_{\sigma,N}(G) \leq \bar{\bar{d}}_{\sigma,N}(G)
\\
\Leftrightarrow
&c_{\sigma,N}(G) \leq \left(\sup_{t \in \mathbb{Z}} \left\| l_t \right\|_{\sigma_0 \infty}\right)^{-1} \rho \quad \text{ (By $(\ref{c_N})$ and $(\ref{bar_bar_d})$)}
\\
\Leftrightarrow
&c_{\sigma,N}(G) \leq \left( \left\| l_t \right\|_{\sigma_0 \infty,t}\right)^{-1} \rho,\forall t \in \mathbb{Z}
\\
\Rightarrow
&c_{\sigma,\sigma_0}(G,t) \leq \left( \left\| l_t \right\|_{\sigma_0 \infty,t}\right)^{-1} \rho,\forall t \in \mathbb{Z} \quad \text{ (by Lemma \ref{c_Nlemma})}
\end{align*}
and hence $(\ref{CS_specical_c1})$ holds. Therefore, by Lemma \ref{stable_xdelayed2x_CS_special}, $\|\Sigma\|_{\infty,t} \leq \bar{c}$ holds. Hence, the corollary is proved. \hfill $\blacksquare$
%=============================================
\begin{remark}
Corollary \ref{stable_xdelayed2x_CS_special_Nwidth} gives an upper bound on average variation rate of loop function $G$ for closed-loop system $\Sigma$ to be stable with degree $1$ given the frozen-time extension $G_{t}$ is stabilizing for all $t \in \mathbb{Z}$, i.e.  $\sup_{t\in\mathbb{Z}}\|l_t\|_{\sigma_0 \infty} < \infty$ for $t \in \mathbb{Z}$.

\hfill $\square$
\end{remark}
%=============================================
\begin{remark}
	\label{Zames_condtion}
	For the system $\Sigma$ to be $\ell_{\sigma \infty e}$-stable, Zames and Wang's  sufficient condition \cite[inequality $(2.22)$]{Zames_Wang_91} is
	\begin{align}
	\left\|\nabla g_t \right\|_{\sigma \infty} \leq \bar{\bar{d}}_{\sigma,1}(G),\forall t \in \mathbb{Z} \label{Zames_cons_c1} 
	\end{align}
	where
	\vspace{-0.1cm}
	\begin{align}
	\bar{\bar{d}}_{\sigma,1}(G) \triangleq \left(e \ln \left(\frac{\sigma_0}{\sigma}\right)\right) 
	\left(\sup_{t \in \mathbb{Z}} \left\| l_t \right\|_{\sigma_0 \infty}\right)^{-1} \rho \label{bar_bar_d_1}
	\end{align}
	with $\rho \in (0,1)$. In \cite{Zames_Wang_91}, Zames and Wang considered the sensitivity function $\left(I-G\mathcal{T}\right)^{-1}$ and proved $\left\|\left(I-G\mathcal{T}\right)^{-1}\right\|_{\sigma \infty} \leq \frac{\sup_{t \in \mathbb{Z}}\left\|s_{t}\right\|_{\sigma \infty}}{1-\rho}$ if $(\ref{Zames_cons_c1})$ holds. On the other hand, we considered the system $\Sigma=\left(I-G\mathcal{T}\right)^{-1}F$, and proved in Lemma \ref{stable_xdelayed2x_CS_special} that $\left\|\left(I-G\mathcal{T}\right)^{-1}F\right\|_{\infty} \leq \bar{c} = \frac{\|F\|_{\infty} \sup_{t \in \mathbb{Z}}\left\|s_{t}\right\|_{\infty}}{1-\rho}$ if the sufficient condition $(\ref{CS_specical_c1})$ holds. \hfill $\square$
\end{remark}
%=============================================
%On the other hand, the sufficient condition $(\ref{CS_specical_c1})$ only requires for all $t \in \mathbb{Z}$,
%\begin{align}
%\sum_{i=t-N+1}^t\left\|\nabla g_i \right\|_{\sigma \infty} < N\bar{\bar{d}}_{\sigma,N} \label{CS_specical_c2}. 
%\end{align} 
%=============================================
\par In the following lemma, the relation between 
our sufficient condition $(\ref{Thm2C1})$ and and Zames and Wang's \cite{Zames_Wang_91} sufficient condition $(\ref{Zames_cons_c1})$ is discussed.
%=============================================
\begin{lemma} \label{less_conservative}
	Consider $\sigma, \sigma_{0} \in \mathbb{R}_{+}$, where $1 \leq \sigma < \sigma_{0}$, and $\rho \in (\sigma^{-1},1)$. Consider the non-linear feedback system $\Sigma$ in Fig. \ref{I-GTinvF} where $F:\mathbb{\ell}^n_{\sigma \infty e} \mapsto \mathbb{\ell}^m_{\sigma \infty e}$ having $\left\|F\right\|_{\infty} < \infty$ and $G:\mathbb{\ell}^m_{\sigma \infty e} \mapsto \mathbb{\ell}^m_{\sigma \infty e}$ with $n,m \in \mathbb{Z}_+\backslash \{0\}$. Let $s_{t}$ and $l_t$ be the TI frozen-time snapshots of $(I-G_{t}\mathcal{T})^{-1}$ and $(I-G_{t}\mathcal{T})^{-1}G_t\mathcal{T}$ respectively. Then the sufficient condition $(\ref{Thm2C1})$ and $(\ref{Thm3C1})$ hold whenever Zames and Wang's sufficient condition $(\ref{Zames_cons_c1})$ holds, and there exist cases where Zames and Wang's sufficient condition $(\ref{Zames_cons_c1})$ does not hold when the sufficient condition $(\ref{Thm2C1})$ and $(\ref{Thm3C1})$ hold.
\end{lemma} 
\emph{Proof:} Refer Appendix E. \hfill $\blacksquare$
%===========================
\begin{remark} 
	By Lemma \ref{less_conservative}, the system $\Sigma$ is $\ell_{\infty e}-$stable when $G$ varies with periodic large-variation such that for all $q \in \mathbb{Z}$, $\left\|\nabla g_{t}\right\|_{\sigma \infty} \in \left( \bar{\bar{d}}_{\sigma,1},N\bar{\bar{d}}_{\sigma,N}\right] ,\forall t = qN$, and $\left\|\nabla g_{t}\right\|_{\sigma \infty}=0,\forall t \neq qN$. $\hfill \square$
\end{remark}
%===========================
\iffalse \section{A limitation of the circle criterion}
For a feedback system where a linear system is subjected to non-linear feedback, the circle criterion \cite{haddad2011nonlinear} returns a finite or infinite range $\left[k_1, k_2\right]$, where $k_1,k_2 \in \mathbb{R} \cup \infty$ and $k_1 \leq k_2$. If the nonlinearity's gain is within the range for all time, then the system is asymptotically stable. However, if the nonlinearity varies abruptly such that its gain momentarily goes out of the range $\left[k_1, k_2\right]$, then the asymptotic stability of the system cannot be concluded by the circle criterion. In this case, instead of the circle criterion we can use Corollary \ref{notweakstab} to determine the $\ell_{\infty e}$-stability of the system by considering the feedback loop consisting of linear system and nonlinearity as loop function $G$. In Example 3 presented later, we showed that Corollary \ref{notweakstab} concludes the stability of a non-linear feedback system with an abruptly time-varying nonlinearity while the circle criterion failed to conclude the asymptotic stability.\fi
%===========================
\section{Comparison with \cite{Zames_Wang_91}}
Zames and Wang \cite{Zames_Wang_91} proved that
\begin{align}
\left\|H \nabla G\right\|_{(\sigma)}
\leq \sigma^{-1}
\left(e \ln\left(\frac{\sigma_0}{\sigma}\right)\right)^{-1}
\left\|H\right\|_{(\sigma_0)}
d_{\sigma}(G), \label{Prop2.2}
\end{align}
where both $H$ and $G$ are causal and linear with bounded $\left\|H\right\|_{\sigma_0 \infty}$ and $\left\|G\right\|_{\sigma \infty}$. By \cite{Zames_Wang_91}, $\left\|H\right\|_{(\sigma_0)} = \sup_{\tau \in \mathbb{Z}}\left\|h_{\tau}\right\|_{\sigma_0 \infty}$. By Remark \ref{dK}, $d_{\sigma}(G)=\sigma\bar{d}_{\sigma,N}(G)$ when $N=1$. By \cite{Zames_Wang_91} and Definition \ref{snapshot}, $\left\|H \nabla G\right\|_{(\sigma)} = \sup_{t \in \mathbb{Z}} \left\|h_t \nabla G_t\right\|_{\sigma \infty,t}$. Therefore $(\ref{Prop2.2})$ is a special case of our Lemma \ref{ZamesLemmaNwidth} when the worst-case variation rate of $G$ is bounded, and $H$ and $G$ are causal, stable, and linear. Therefore, Lemma \ref{ZamesLemmaNwidth} generalizes $(\ref{Prop2.2})$ by considering non-linear and unstable $H$ and $G$ and by relaxing the assumption that the worst-case variation rate of $G$ is bounded.
%-----------------
\par Theorem \ref{stable_xdelayed2x_CS} generalizes Zames and Wang's sufficient condition $(\ref{Zames_cons_c1})$ by relaxing assumptions that $F$ and $G$ in Fig. \ref{feedbacksys} are linear and the worst-case variation rate of $G$ is bounded. Theorem \ref{stable_xdelayed2x_CS} allows unstable $\left(I-G\mathcal{T}\right)^{-1}G\mathcal{T}$ such that its frozen-time snapshot is not necessarily $\ell_{\sigma_0 \infty e}$-stable for all time. Furthermore, Theorem \ref{stable_xdelayed2x_CS} considers weakly ${\ell}_{\infty e}$-stability of the system $\Sigma$ at given time sequence $\left\{t_i\right\}$, which generalizes Zames and Wang's sufficient condition $(\ref{Zames_cons_c1})$ where ${\ell}_{\sigma \infty e}-$stability of the system $\Sigma$ is considered for all time. By Lemma \ref{less_conservative}, Zames and Wang's sufficient condition $(\ref{Zames_cons_c1})$ is a special case of Theorem \ref{stable_xdelayed2x_CS}.
%=====================================================
\section{Bound on Plant Time-variation Rate for Adaptive Control}
\begin{figure}[t]
	\centering
	\includegraphics[width=.4\textwidth]{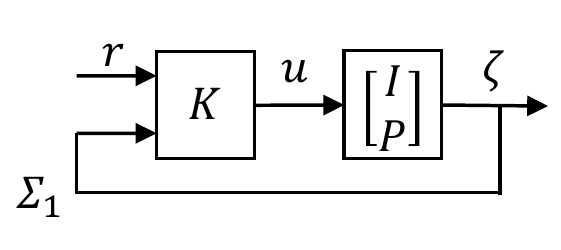}
	\caption{The adaptive switched system $\Sigma_1$ considered in \cite{patil2016nonlinear}.}
	\vspace{0cm}
	\label{Sagar_sys}
\end{figure}

\begin{figure}[t]
	\centering
	\includegraphics[width=.475\textwidth]{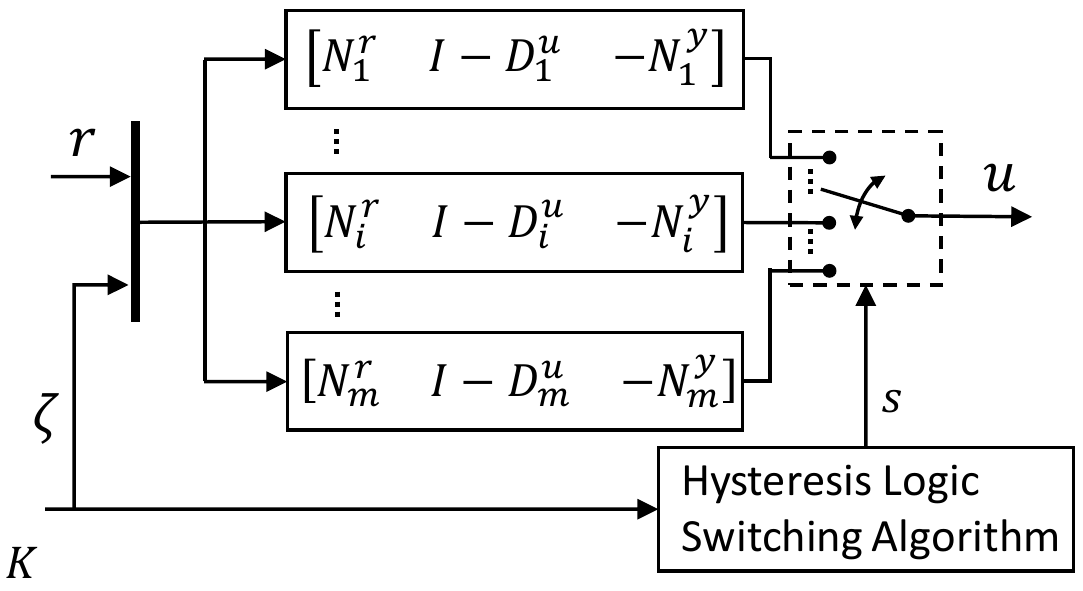}
	\caption{The nonlinear adaptive controller $K$ implemented in the adaptive switched system $\Sigma_1$.}
	\vspace{0cm}
	\label{Sagar_K}
\end{figure}

%----Sagar's ti is the ti-1 in the present paper.
An interesting question in adaptive control is how much plant time-variation rate can be tolerated. In this section, we answer this question with the help of Corollary \ref{stable_xdelayed2x_CS_special_Nwidth} to derive an upper bound on allowed average plant time-variation rates in adaptive control framework. We consider the adaptive switching system developed in \cite{patil2016nonlinear} as follows. For an unknown \emph{slowly time-varying} nonlinear plant $P$, the paper \cite{patil2016nonlinear} proposes an algorithm that returns a stabilizing adaptive switching controller $K$ of the form shown in Fig. \ref{Sagar_K} such that the resultant closed-loop adaptive system $\Sigma_1 \left(K,P\right)$ of the form shown in Fig. \ref{Sagar_sys} with resetting is exponentially stable and has bounded $\ell_{\infty e}$-norm subject to the assumption that the adaptive control problem is \emph{feasible} in the sense there always exist at least one candidate controller capable of stabilizing the \emph{slowly time-varying} plant $P$, where $\underline{M}\triangleq \left\{1,2,\dots,m \right\}$ is the set of candidate controllers' indices. The importance of our Corollary \ref{stable_xdelayed2x_CS_special_Nwidth} is that it not only confirms that the system will remain stable in the presence of slow and/or infrequent large plant time-variation but also gives a quantitative bound on the amount of tolerable average rate of plant time-variation, provided that the frozen-time adaptive problem for the frozen time plants $P_t$ are feasible for all $t\in\mathbb{Z}_+$ and the average variation rate of the frozen-time snapshots of the open-loop system is small enough.

\par The adaptive switching system $\Sigma_1$ can be converted to the generic feedback system in Fig. \ref{feedbacksys} by letting $F = \begin{bmatrix} I \ P \end{bmatrix}' N_{s}^r$, $x = \zeta$, and $G\mathcal{T} = \begin{bmatrix} I \ P \end{bmatrix}' \begin{bmatrix}I-D_{s}^u & -N_{s}^y \end{bmatrix}$. Let $l_t$ be the frozen-time snapshot of $\left(I-G_t\mathcal{T}\right)^{-1}G_t\mathcal{T}$ for all $t\in \mathbb{Z}$. The paper \cite{patil2016nonlinear} has proved that the proposed nonlinear adaptive controller $K$ achieves $\ell_{\lambda \infty e}$-stability for all $P_t, t \in \mathbb{Z}$ with some degree $\lambda>1$ which is subject to the feasibility assumption \cite[Assumption A2]{patil2016nonlinear}, and thus we have $\left\|l_t\right\|_{\lambda \infty}$ bounded for all $t \in \mathbb{Z}$. We assume that the $N$-width average time-variation of plant $P$ is bounded by $\bar{d}_{\sigma,N}(P) $ for some $N\in\mathbb{Z}_+ \backslash \{0\}$ and $\sigma \in \left(1,\lambda\right)$. Then by Lemma \ref{d_sigma_N_of_prod}, we have
\begin{align} \bar{d}_{\sigma,N}(G) \leq &\max_{i \in \underbar{M}}\left\|\begin{bmatrix}I-D_{i}^u & -N_{i}^y \end{bmatrix}\right\|_{\sigma \infty} \bar{d}_{\sigma,N}(P) \label{d_bar_of_p}
\end{align}
where $\max_{i \in M}\|[I-D_i^u \ -N_i^y]\|_{\sigma\infty}$ is finite according to controller realization in \cite{patil2016nonlinear}. According to Corollary \ref{stable_xdelayed2x_CS_special_Nwidth} and $(\ref{d_bar_of_p})$, if for some $\rho \in \left(\sigma^{-1},1\right)$ the term $\bar{d}_{\sigma, N}(P)$ satisfies
\begin{align}
\bar{d}_{\sigma,N}(P)) \leq  
&\left(\max_{i \in \underbar{M}}\left\|\begin{bmatrix}I-D_{i}^u & -N_{i}^y \end{bmatrix}\right\|_{\sigma \infty}\right)^{-1}
\left(\frac{\lambda}{\sigma}\right)^{1-N} \nonumber
\\ 
&\quad e \ln \left(\frac{\lambda}{\sigma}\right)
\left(\sup_{t \in \mathbb{Z}} \left\| l_t \right\|_{\lambda \infty}\right)^{-1} 
\rho, \label{d_bar_of_p_condition}
\end{align}
%=============
%=============
then $\ell_{\infty e}$-stability of the system $\Sigma_1$ is preserved. Therefore, as long as the $N$-width average variation rate of plant $P$ does not violet the inequality $(\ref{d_bar_of_p_condition})$, the nonlinear adaptive controller $K$ developed in \cite{patil2016nonlinear} preserves $\ell_{\infty e}$-stability of the adaptive switching system $\Sigma_1$.
%=====================================================
\section{Simulation}
\par In this section, \textsc{Matlab} simulations are presented. The Examples 1 is demonstrated to support the Corollary \ref{notweakstab}. And the Example 2 shows a case where Zames and Wang's condition $(\ref{Zames_cons_c1})$ does not hold while our condition $(\ref{Thm3C1})$ holds and concludes $\ell_{\infty e}$-stability of a system.\iffalse The \textsc{MATLAB} simulation Example 3 shows a case where the circle criterion does not conclude $\ell_{\infty e}$-stability of a non-linear feedback system while \textcolor{red}{our condition $(\ref{Thm3C1})$ does}.\fi
%====================================
%==================EXAMPLE 1
%\\ \begin{center} \bfseries{Example 1} \normalfont \end{center}
\par \bfseries{\textit{Example 1:}} \normalfont Consider the system $\Sigma$ in Fig. \ref{feedbacksys}. Let $F$ be an identity matrix. Let the persistently destabilizing loop function $G$ be equal to a time-varying non-linear system $\Phi H_{t} $ such that $\left(Gx\right)(t)=\Phi H_t x(t)$ for all $t \in \mathbb{Z}$ where $x \in \ell^2_{\sigma \infty e}$. The system $\Phi$ with input $[v_1 \ v_2]'$ and output $[w_1 \ w_2]'$ is a dead-zone operator such that for all $i \in \{1,2\}$,
%==================
\begin{figure}[ht]
	\centering
	\includegraphics[width=.5\textwidth]{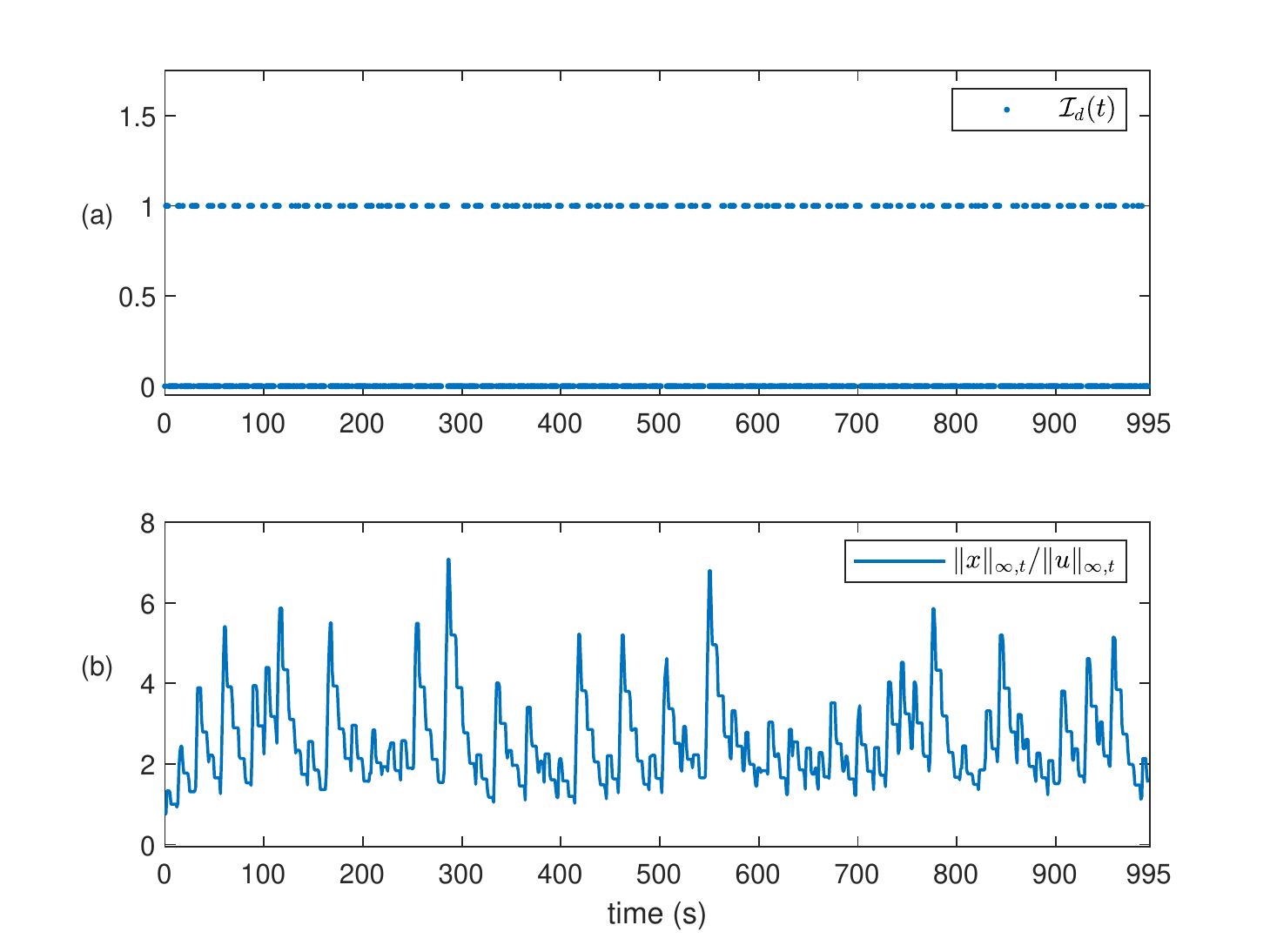}
	\caption{(a) Frozen-time extensions $G_t$ are destabilizing whenever $\mathcal{I}_d(t)=1$. (b) Gain of system $\Sigma$.}
	\vspace{0cm}
	\label{ex1_plot_1}
\end{figure}
%===================
\begin{align}
w_i=
\begin{cases}
v_i-0.5, &\text{ if } v_i \geq 0.5,
\\
0,&\text{ if } v_i \in(-0.5,0.5),
\\
v_i+0.5,&\text{ if } v_i \leq 0.5,
\end{cases} \label{deadzone}
\end{align} 
and the system $H_t$ is time-varying such that (i) $\left(H_t x\right)(t)=A_tx(t) + B_t x(t-1)$ where $\lambda_{\text{max}}(B_t) < 1$ for all $t\in[0,995]$, (ii) $\lambda_{\text{max}}(A_t) < 1$ whenever the function $\mathcal{I}_d(t) = 1$, and $\lambda_{\text{max}}(A_t) > 1$ whenever $\mathcal{I}_d(t) = 0$ for all $t\in[0,995]$ as shown in Fig. \ref{ex1_plot_1}(a), (iii) $H_t$ is destabilizing whenever $\mathcal{I}_d(t) = 1$ and $H_t$ is stabilizing whenever $\mathcal{I}_d(t) = 0$, and (iv) $H_i \neq H_j, \forall i,j \in [0,995]$ and $i \neq j$.
\par We simulated the above system $\Sigma$ in \textsc{MATLAB} with zero initial conditions and $u(t)=2\exp\left(\frac{t}{20}\right)\cos\left(\frac{t}{2}\right)\begin{bmatrix}1 & 1\end{bmatrix}'$ for $t = 0:995$. We considered $\sigma=1.2$, $\sigma_{0}=1.4$, and $\rho=0.94$. The simulation results are shown in Fig. \ref{ex1_plot_1}(b).
%=================
\begin{figure}[t]
	\centering
	\includegraphics[width=0.5\textwidth]{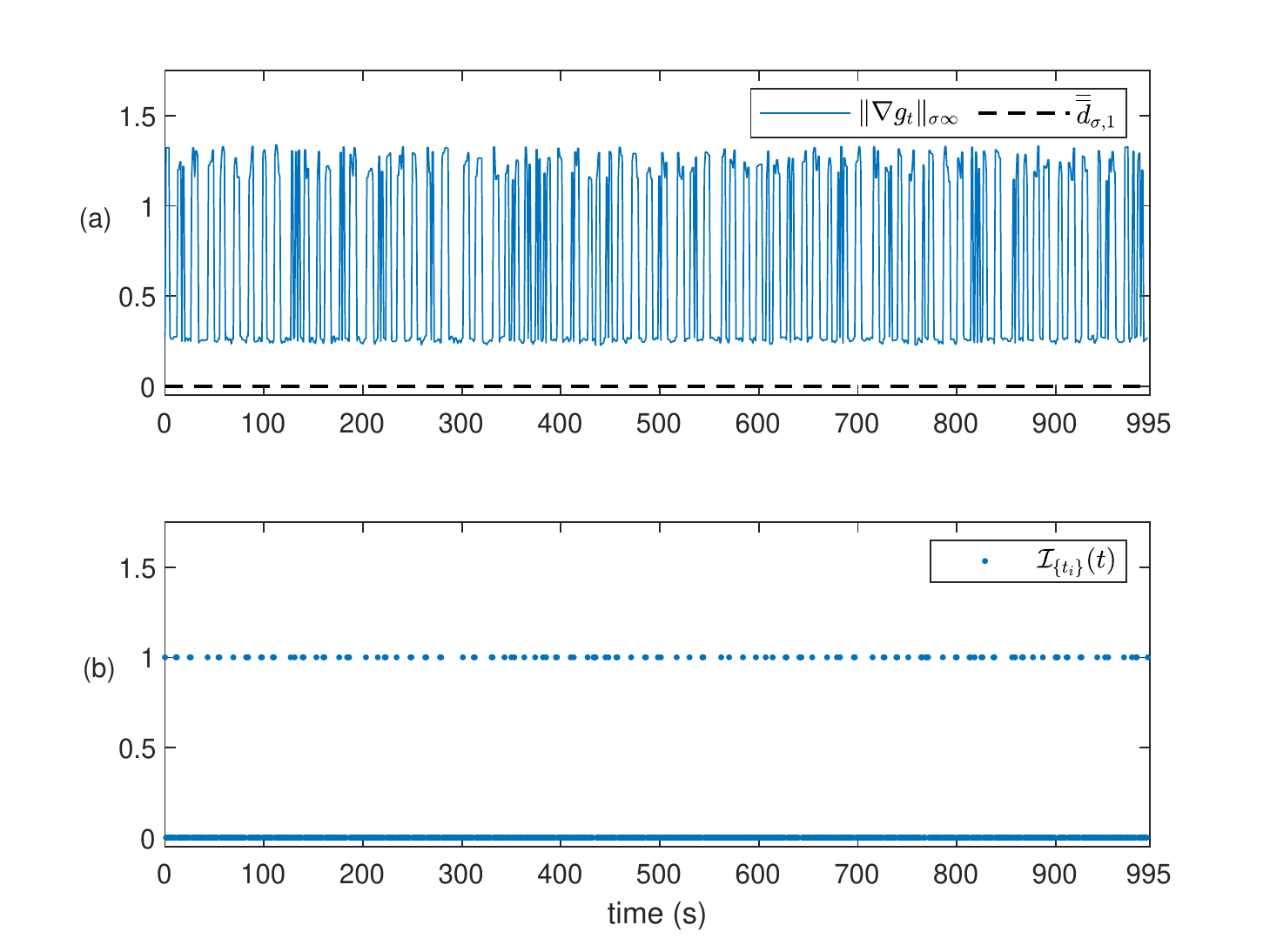}
	\caption {(a) Persistent and abrupt time-variations in $G$. (b) Time sequence $\left\{t_i\right\}=\left\{\mathcal{I}_{t_i}(t)=1,t \in [0,995]\right\}$.}
	\vspace{0cm}
	\label{ex1_plot_2}
\end{figure}
%=================
\par
\begin{figure}[ht]
	\vspace{0cm}
	\centering
	\includegraphics[width=0.5\textwidth]{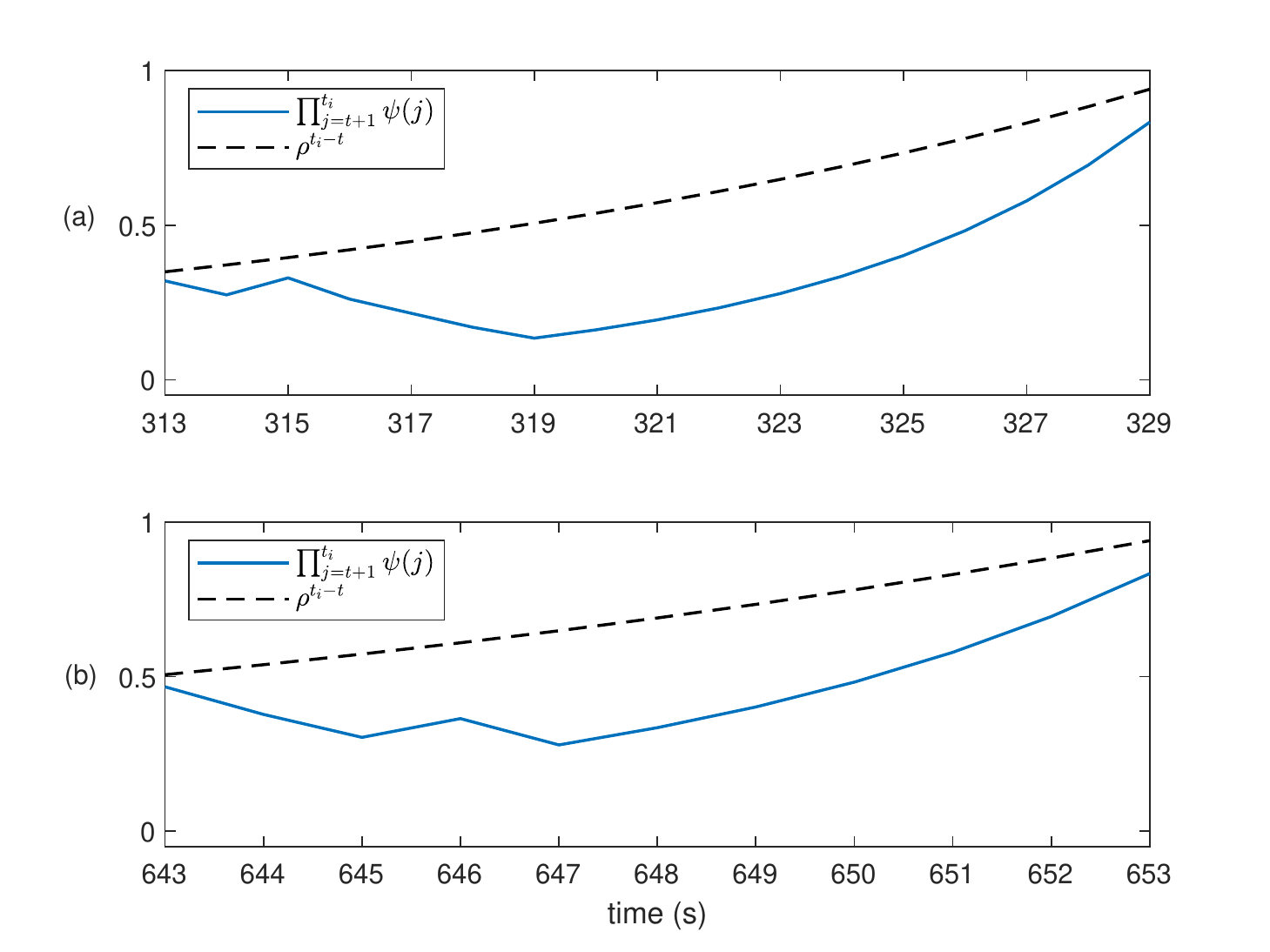}
	\caption {(a) Condition $(\ref{Thm3C1})$ holds for $\left(t_{i-1},t_i\right)=\left(313,330\right)$. (b) Condition $(\ref{Thm3C1})$ holds for $\left(t_{i-1},t_i\right)=\left(643,654\right)$.}
	\vspace{0cm}
	\label{ex1_plot_3}
\end{figure}

 Persistent and abrupt time-variations in the loop function $G$ are shown in Fig. \ref{ex1_plot_2}(a). First we computed the terms $c_{\sigma,\sigma_0}(G)(t)$ and $\hat{\psi}(t)$ for all $t \in [0,995]$ by $(\ref{c_sigma})$ and $(\ref{globalpsi})$ respectively. We choose a time sequence $\left\{t_i\right\}$ shown in Fig. \ref{ex1_plot_2}(b). Then by the sufficient condition $(\ref{Thm3C1})$, the $\ell_{\infty e}$-stability of the system $\Sigma$ is preserved since (i) $\left\|x \right\|_{\infty,0}=0$ because of the zero initial condition and (ii) condition $(\ref{Thm3C1})$ holds for all $t_i \in \left\{t_i\right\}$. For example, Fig. \ref{ex1_plot_3}(a) and (b) show the condition $(\ref{Thm3C1})$ holds for $\left(t_{i-1},t_i\right)=\left(313,330\right)$ and $\left(t_{i-1},t_i\right)=\left(643,654\right)$ respectively. By $(\ref{globalc})$, we compute $\hat{c}=79129$. Therefore, by Corollary \ref{notweakstab}, $\|\Sigma\|_{\infty,t} \leq \hat{c}$ which can be verified in Fig. \ref{ex1_plot_1}(b) where $\frac{\|x\|_{\infty,t}}{\|u\|_{\infty,t}} \leq \hat{c}$ for all $t \in \left[0,995\right]$.
\par On the other hand, since $H(t)$ is destabilizing whenever $\mathcal{I}_{d}(t)=1$ for all $t\in [0,995]$, the frozen-time snapshot $l_t$ of $\left(I-G_t\mathcal{T}\right)^{-1}G_t\mathcal{T}$ is unstable, and $\sup_{t \in\mathbb{Z}}\left\|l_t\right\|_{\sigma_0 \infty} = \infty$ according to Definition \ref{sigmapsysnorm}. Zames and Wang's sufficient condition $(\ref{Zames_cons_c1})$ for the system $\Sigma$ to be $\ell_{\sigma \infty e}$-stable is $\left\|g_t\right\|_{\sigma \infty} \leq \bar{\bar{d}}_{\sigma,1}(G)$ for all $t \in \mathbb{Z}$. By $(\ref{bar_bar_d_1})$, we computed $\bar{\bar{d}}_{\sigma,1}(G)=0$. But, $\left\|g_t\right\|_{\sigma \infty}>0$ for all $t \in [0,995]$ as shown in Fig. \ref{ex1_plot_2}(a). Therefore, Zames and Wang's sufficient condition $(\ref{Zames_cons_c1})$ does not hold for all $t \in [0,995]$ as shown in Fig. \ref{ex1_plot_2}(a) and so it does not conclude that the system $\Sigma$ is $\ell_{\sigma \infty e}$-stable. This proves for this example that our sufficient condition $(\ref{Thm3C1})$ is less conservative than Zames and Wang's sufficient condition $(\ref{Zames_cons_c1})$, i.e. the condition $(\ref{Thm3C1})$ holds while the condition $(\ref{Zames_cons_c1})$ does not hold.
%==================EXAMPLE 2
\par \bfseries{\textit{Example 2:}} \normalfont Consider the system $\Sigma$ in Fig. \ref{feedbacksys}. Let $F$ be an identity matrix. Let the loop function $G$ be equal to the system $H_{t}$ such that $\left(Gx\right)(t)=H_t x(t)$ for all $t \in \mathbb{Z}$ where $x \in \ell^2_{\sigma \infty e}$. And the system  $H_t$ is a time-varying $2\times 2$ real-matrix such that (i)$\left|\lambda_{\text{max}}\left(H_{t}\right)\right| < 1$ for all $t \in \mathbb{Z}_{+}$ and (ii) $H_i \neq H_j, \forall i,j \in \mathbb{Z}_{+}$ and $i \neq j$. The frozen-time snapshot $l_t$ of $\left(I-G_t\mathcal{T}\right)^{-1}G_t\mathcal{T}$ has $\sup_{t \in \mathbb{Z}_{+}}\|l_t\|_{\sigma_0 \infty,t}=4.8839$ with considered $G$.
\par
\begin{figure}[ht]
	\centering
	\includegraphics[width=.5\textwidth]{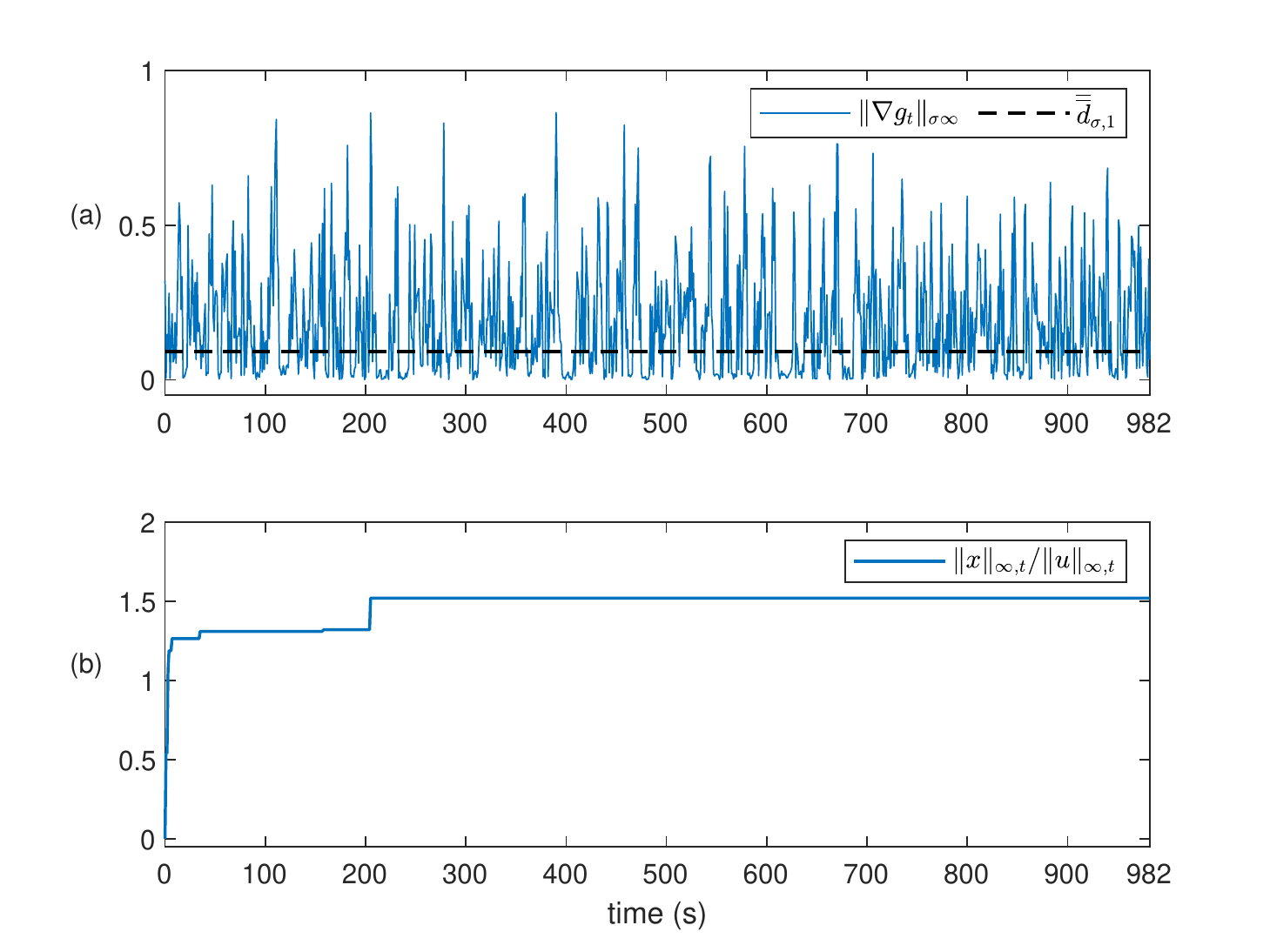}
	\caption{(a) Persistently varying frozen-time extensions $G_t$. (b) Gain of system $\Sigma$.}
	\vspace{-0.1cm}
	\label{ex2_plot_1}
\end{figure}

 We simulated the above system $\Sigma$ in \textsc{MATLAB} with zero initial conditions and $u=\left[\cos(t/2) \ \cos(t/2)\right]'$ for $t=0:982$. We considered $\sigma=1.2$, $\sigma_{0}=1.44$, and $\rho=0.9$. The simulation results are shown in Fig. \ref{ex2_plot_1}.
\par
\begin{figure}[t]
	\centering
	\includegraphics[width=.5\textwidth]{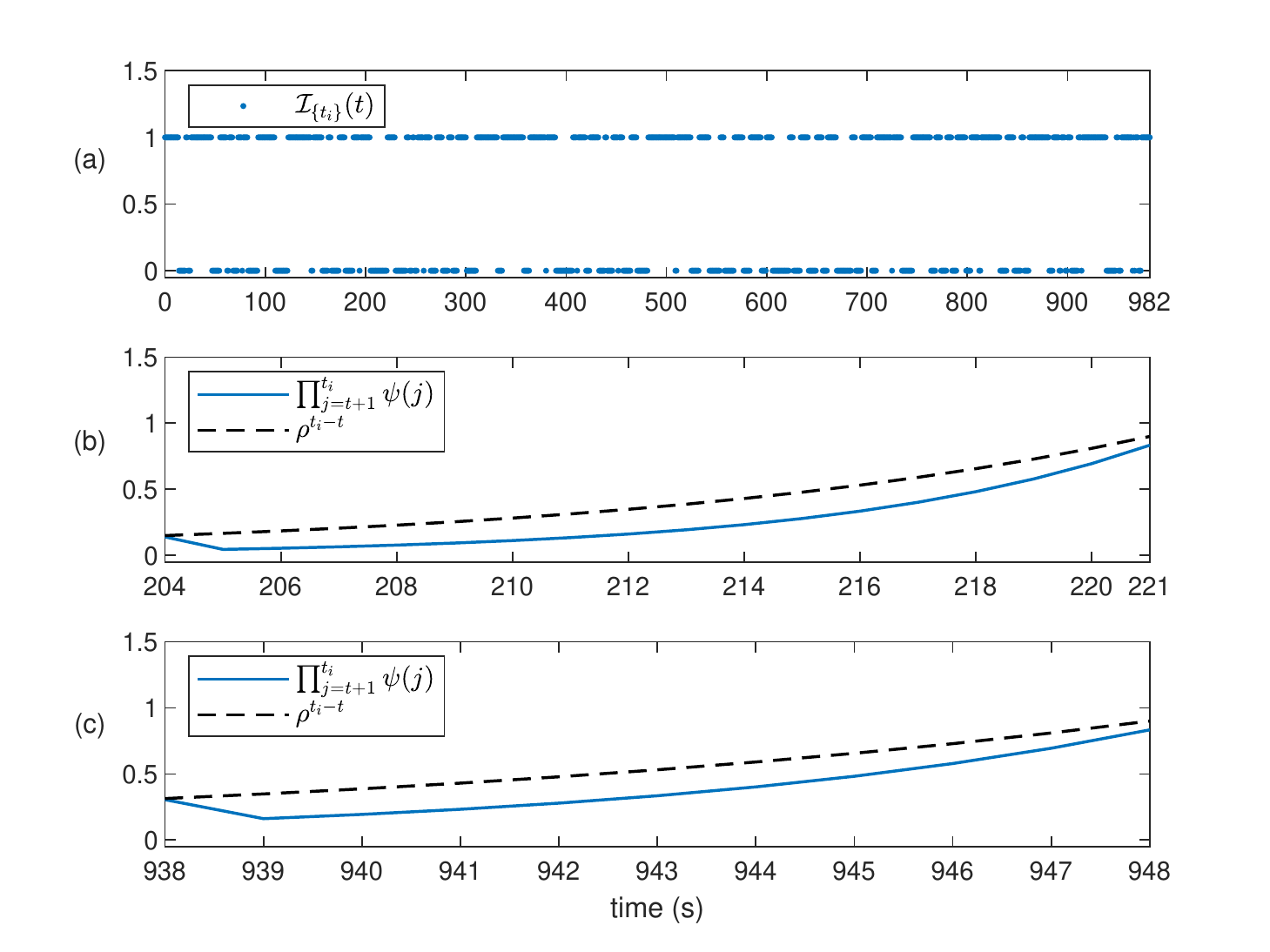}
	\caption{(a) Time sequence $\left\{t_i\right\}=\left\{\mathcal{I}_{t_i}(t)=1,t\in[0,982]\right\}$. (b) Condition $(\ref{Thm3C1})$ holds for $\left(t_{i-1},t_i\right)=\left(204,222\right)$. (c) Condition $(\ref{Thm3C1})$ holds for $\left(t_{i-1},t_i\right)=\left(938,949\right)$.}
	\vspace{-0.1cm}
	\label{ex2_plot_2}
\end{figure}

 Fig. \ref{ex2_plot_1}(a) shows persistent and abrupt time-variations in $G$. First we computed the terms $c_{\sigma,\sigma_0}(G)(t)$ and $\hat{\psi}(t)$ for all $t \in [0,982]$ by $(\ref{c_sigma})$ and $(\ref{globalpsi})$ respectively. We choose a time sequence $\left\{t_i\right\}$ as shown in Fig. \ref{ex2_plot_2}(a). Then by the sufficient condition $(\ref{Thm3C1})$, the $\ell_{\infty e}$-stability of the system $\Sigma$ is preserved since (i) $\left\|x \right\|_{\infty,0}=0$ because of the zero initial condition and (ii) condition $(\ref{Thm3C1})$ holds for all $t_i \in \left\{t_i\right\}$. For example, Fig. \ref{ex2_plot_2}(b) and (c) show the condition $(\ref{Thm3C1})$ holds for $\left(t_{i-1},t_i\right)=\left(204,222\right)$ and $\left(t_{i-1},t_i\right)=\left(938,949\right)$ respectively. By $(\ref{globalc})$, we computed $\hat{c}=6266$. Therefore, by Corollary \ref{notweakstab}, $\|\Sigma\|_{\infty,t} \leq \hat{c}$ which can be verified in Fig. \ref{ex2_plot_1}(b) where and $\frac{\|x\|_{\infty,t}}{\|u\|_{\infty,t}} \leq \hat{c}$ for all $t \in \left[0,982\right]$.
\par On the other hand, by $(\ref{bar_bar_d_1})$, we computed $\bar{\bar{d}}_{\sigma,1}(G)=0.0913$. Since $\left\|\nabla g_t\right\|_{\sigma \infty} > \bar{\bar{d}}_{\sigma,1}(G)$ for most of the time as shown in Fig. \ref{ex2_plot_1}(a), the condition $(\ref{Zames_cons_c1})$ does not hold. Therefore, Zames and Wang's sufficient condition $(\ref{Zames_cons_c1})$ does not conclude that the ${\ell}_{\infty e}$-stability of the simulated system is preserved. This proves for this example that our sufficient condition $(\ref{Thm3C1})$ is less conservative than Zames and Wang's sufficient condition $(\ref{Zames_cons_c1})$, i.e. the condition $(\ref{Thm3C1})$ holds while the condition $(\ref{Zames_cons_c1})$ does not hold.

\section{Conclusion}
\par In this article, the input-output stability of a general time-varying MIMO non-linear feedback system has been investigated by generalizing the results in \cite{Zames_Wang_91}. A general sufficient condition to preserve stability of the feedback system has been derived by relaxing three assumptions \cite{Zames_Wang_91} on the adaptive feedback loop function that (i) it is linear, (ii) its frozen-time snapshot is stabilizing all the time, and (iii) variation between its adjacent frozen-time snapshots is bounded. The sufficient condition gives a tolerable limit on average time-variation rate of the adaptive feedback loop function of a MIMO non-linear adaptive switching system to preserve its $\ell_{\infty e}$-stability. 
\par Our sufficient condition is less conservative compared to the sufficient condition in \cite{Zames_Wang_91}. Whenever the condition \cite{Zames_Wang_91} holds, our condition holds as well. In case when the adaptive feedback loop function has infrequent large time-variations, our condition holds but the condition \cite{Zames_Wang_91} does not hold. Therefore, our condition is more practical to conclude stability of adaptive switching systems that are inherently non-linear and subject to infrequent large variations possibly due to unexpected component failures.
\iffalse \par Additionally, because the present sufficient condition considers average of time-variation rates of loop functions instead of worst-case time-variation rates between two adjacent time instants, the present sufficient condition can be used to conclude stability of non-linear feedback systems that cannot be concluded by the circle criterion when the loop functions have large and infrequent time-variations.\fi
%------------------------------------------------- Lemma 4.1 proof ------------------------------------------------------------------------------------
\begin{center} \textit\bfseries{Appendix A \\ Proof of Lemma \ref{ZamesLemma}} \end{center}
\par Consider the system $h_t \nabla G_t$ with input $u \in \mathbb{\ell}^n_{\sigma \infty e}$ and output $y \in \mathbb{R}^m$ where $n,m \in \mathbb{Z}_+\backslash \{0\}$. Then by Lemma \ref{delAndela}, $\forall t \in \mathbb{Z}$ we have
\begin{align*}
y(t)&=
h_t \nabla G_t u
=
h_t
\begin{bmatrix}
\vdots \\ \left(\sum_{i=t-1}^t \nabla g_i\right) \mathcal{T}^2u \\ \nabla g_t \mathcal{T}u \\ 0
\end{bmatrix}.
\end{align*}
Next,
\begin{align}
&|y(t)| \nonumber
\\
%------------
\leq&
\left\|h_t\right\|_{\sigma_0 \infty,t}
\left\| 
\begin{bmatrix}
\vdots \\ \left(\sum_{i=t-1}^t \nabla g_i\right) \mathcal{T}^2u \\ \nabla g_t \mathcal{T}u \\ 0
\end{bmatrix} \right\|_{\sigma_0 \infty , t} \nonumber
%-------------
\\
=&
\left\|h_t\right\|_{\sigma_0 \infty,t}
\left\| 
\begin{bmatrix}
\vdots \\ \sigma_0^{-2} \left(\sum_{i=t-1}^t \nabla g_i\right) \mathcal{T}^2u \\ \sigma_0^{-1} \nabla g_t \mathcal{T}u \\ 
0
\end{bmatrix} \right\|_{\infty} \nonumber
%------------
%\\
%\leq & \left\|h_t\right\|_{\sigma_0 \infty,t}
%\left\|
%\begin{bmatrix}
%\vdots \\ \frac{{\sigma}^2}{{\sigma_0}^2}\left\|k_{t-2}-k_t\right\|_{\sigma %\infty,t-2}\left\|u_{t-2}\right\|_{\sigma \infty,t} \\ 
%\frac{\sigma}{\sigma_0} \left\|k_{t-1}-k_t\right\|_{\sigma %\infty,t-1}\left\|u_{t-1}\right\|_{\sigma \infty,t}  \\ 
%0
%\end{bmatrix}\right\|_{\infty} \nonumber
%\end{align}
%------------
%\begin{align}
\\
\leq& \left\|h_t\right\|_{\sigma_0 \infty,t}
\left\|
\begin{bmatrix}
\vdots \\ \frac{{\sigma}^2}{{\sigma_0}^2} \sum_{i=t-1}^t \left\| \nabla g_i \right\|_{\sigma \infty}\\ 
\frac{\sigma}{\sigma_0}\left\|\nabla g_t\right\|_{\sigma \infty}  \\ 
0
\end{bmatrix}\right\|_{\infty}\left\|u\right\|_{\sigma \infty,t} \label{ZamesLemmaP1}
%------------
\\
\leq& 
\left\|h_t\right\|_{\sigma_0 \infty,t}
\sup_{i \geq 1}
\left[\left(\frac{\sigma}{\sigma_0}\right)^i
\sum_{q=t-i+1}^{t}\left\|\nabla g_{q}\right\|_{\sigma \infty} \right]
\left\|u\right\|_{\sigma \infty,t} \label{ZamesLemmaP2}
\end{align}
where $(\ref{ZamesLemmaP1})$ is by Lemma \ref{pnormInq}, and $(\ref{ZamesLemmaP2})$ is by the definition of ${\ell}_\infty$-norm. Hence, the claim is proved by $(\ref{ZamesLemmaP2})$ and Definition \ref{sigmapsysnorm}. \hfill $\blacksquare$
%------------------------------------------------- Lemma 4.2 proof ------------------------------------------------------------------------------------
\begin{center} \textit\bfseries{Appendix B \\ Proof of Lemma \ref{c_Nlemma}} \end{center}
\par Consider $i \in \mathbb{Z}_+ \setminus \{0\}$. Since $\forall t \in \mathbb{Z}$, $\exists j \in \mathbb{Z}_+ \setminus \{0\}$ such that $ t-i+1 \in [t-jN+1,t-(j-1)N]$ and $i \in \left[(j-1)N+1,jN\right]$, thus by Definition \ref{Nwidth},
\vspace{2cm}
\begin{align}
&\left(\frac{\sigma}{\sigma_0}\right)^i
\sum_{q=t-i+1}^{t}\left\|\nabla g_{q}\right\|_{\sigma \infty} \nonumber
\\
\leq & 
\left(\frac{\sigma}{\sigma_0}\right)^{(j-1)N+1}
\sum_{l=1}^j\sum_{q=t-lN+1}^{t-(l-1)N} \left\|\nabla g_q \right\|_{\sigma \infty} \nonumber
%---------------
\\
\leq  
&\left(\frac{\sigma}{\sigma_0}\right)^{1-N} 
\bar{d}_{\sigma,N}(G)  
\left(\frac{\sigma}{\sigma_0}\right)^{jN} jN \nonumber
%---------------
\\
%\end{align}
%\begin{align}
\leq 
& \underbrace{
	\left(\frac{\sigma}{\sigma_0}\right)^{1-N} 
	\bar{d}_{\sigma,N}(G)
	\left(e \ln \left(\frac{\sigma_0}{\sigma}\right)\right)^{-1}}_{c_{\sigma,N}(G)} \label{c_NlemmaP1}
\end{align}
where $(\ref{c_NlemmaP1})$ is due to $\sup_{x \geq 0} x y^{-x} \leq \left(e \ln(y) \right)^{-1},\forall y > 1$ \cite{Zames_Wang_91}. Hence, the claim is proved by $(\ref{c_NlemmaP1})$ and $(\ref{c_sigma})$. \hfill $\blacksquare$
%------------------------------------------------- Theorem 4.1 proof ------------------------------------------------------------------------------------
\begin{center} \textit\bfseries{Appendix C \\ Proof of Theorem \ref{stable_xdelayed2x_CS}} \end{center}
\begin{figure}[ht]
	\centering
	\includegraphics[width=\linewidth]{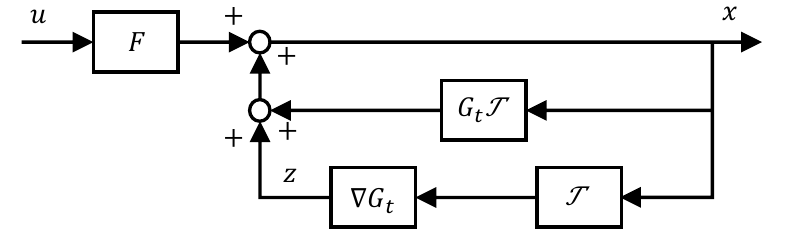}
	\caption{The system $\Sigma$ in terms of $G_t$ and $\nabla G_t$.}
	\vspace{0cm}
	\label{xdelayed2xF1}
\end{figure}

\par Let $t \in \left[t_{i-1}+1,t_i\right]$ for some $i \in \mathbb{Z}$. By Definition \ref{snapshot}, $G\mathcal{T}=G_t\mathcal{T}+\nabla G_t \mathcal{T}$, and so the system $\Sigma$ can be depicted as in Fig. \ref{xdelayed2xF1}.

\par Let $I$ be an identity operator such that $Ix=x$. Then according to Fig. \ref{xdelayed2xF1}, $x = Fu + G_{t}\mathcal{T}x + \nabla G_{t}\mathcal{T}x$  \footnote{According to Definitions \ref{signal} and \ref{system}, $F$, $G_t$, and $\nabla G_t$ are not necessarily real matrices, and hence $\Sigma$ is not necessarily a system in the state-space representation. In a special case where $F$, $G_t$, and $\nabla G_t$ are memory-less systems and thus can be represented as real matrices, $\Sigma$ is a system expressed in the state-space representation.} and thus
\begin{align}
x&=\left(I-G_t\mathcal{T}\right)^{-1}Fu+\left(I-G_t\mathcal{T}\right)^{-1}\nabla G_t \mathcal{T}x \nonumber
\\
&=\left(I-G_t\mathcal{T}\right)^{-1}Fu+\left[I+\left(I-G_t\mathcal{T}\right)^{-1}G_t\mathcal{T}\right]\nabla G_t \mathcal{T}x \nonumber
\\
&=\left(I-G_t\mathcal{T}\right)^{-1}Fu+I\nabla G_t \mathcal{T}x \nonumber
\\
&\qquad + \left[\left(I-G_t\mathcal{T}\right)^{-1}G_t\mathcal{T}\right]\nabla G_t \mathcal{T}x. \label{Thm2P0}
\end{align}
Let $k_t$ be the frozen-time snapshot of $I\nabla G_t$, then $k_t \mathcal{T}x=0$ by the fact that $I$ is memory-less and by Definition \ref{snapshot}. Next, since $s_t$ and $l_t$ are the frozen-time snapshots of $\left(I-G_t\mathcal{T}\right)^{-1}$ and $\left(I-G_t\mathcal{T}\right)^{-1}G_t\mathcal{T}$ respectively, then by $(\ref{Thm2P0})$,
\begin{align}
x(t)&=s_tFu+k_t \mathcal{T}x + l_t\nabla G_t \mathcal{T}x=s_tFu+l_t\nabla G_t \mathcal{T}x, \nonumber
\\
|x(t)|& 
\leq 
\left\|s_t\right\|_{\infty,t} 
\left\|F\right\|_{\infty,t}
\left\|u\right\|_{\infty,t} 
+
\left\|l_t \nabla G_t\right\|_{\sigma \infty,t}
\left\|\mathcal{T}x\right\|_{\sigma \infty,t} \nonumber
\\
&\leq
\left\|s_t\right\|_{\infty,t} 
\left\|F\right\|_{\infty}
\left\|u\right\|_{\infty,t} 
+
\left\|l_t \nabla G_t\right\|_{\sigma \infty,t}
\left\|x\right\|_{\sigma \infty,t-1} \nonumber
\\
& \leq
\left\|s_t\right\|_{\infty,t} 
\left\|F\right\|_{\infty}
\left\|u\right\|_{\infty,t} \nonumber
\\
& \qquad
+
\left\|l_t\right\|_{\sigma_0 \infty,t}
c_{\sigma,\sigma_0}(G,t)
\left\|x\right\|_{\sigma \infty,t-1}, \label{Thm2P1}
\end{align}
where $(\ref{Thm2P1})$ is by Lemma $\ref{ZamesLemma}$.

\par Additionally, by Fig. \ref{feedbacksys}, $\left\|s_t\right\|_{\infty,t}\geq 1$ for $t \in \mathbb{Z}$ and Definition \ref{snapshot}, we have
\begin{align}
x(t) &= f_tu+g_t \mathcal{T}x, \nonumber
\\
|x(t)|& 
\leq 
\left\|F\right\|_{\infty}
\left\|u\right\|_{\infty,t} 
+
\left\|g_t\right\|_{\sigma \infty,t}
\left\|\mathcal{T}x\right\|_{\sigma \infty,t} \text{(by Lemma \ref{snapshot_norm_bound})} \nonumber
\\
|x(t)|& 
\leq 
\left\|s_t\right\|_{\infty,t} 
\left\|F\right\|_{\infty}
\left\|u\right\|_{\infty,t} 
+
\left\|g_t\right\|_{\sigma \infty,t}
\left\|\mathcal{T}x\right\|_{\sigma \infty,t} \nonumber
\\
&\leq
\left\|s_t\right\|_{\infty,t} 
\left\|F\right\|_{\infty}
\left\|u\right\|_{\infty,t} 
+
\left\|g_t\right\|_{\sigma \infty,t}
\left\|x\right\|_{\sigma \infty,t-1} \label{Thm2P1.5}
\end{align}

\par Next, the property of $\ell_{\sigma\infty}$-semi norm $\left\|x\right\|_{\sigma \infty,t} \leq 
\max\left\{\sigma^{-1}\left\|x\right\|_{\sigma \infty,t-1},\left|x(t)\right|\right\}$ along with $(\ref{Thm2P1})$ and $(\ref{Thm2P1.5})$ are used to get
\begin{align}
&\left\|x\right\|_{\sigma \infty,t} \nonumber
\\
\leq
&\max \big \{ \sigma^{-1} \left\|x\right\|_{\sigma \infty,t-1}, \nonumber
\\
&\,\min\{
\left\|s_t\right\|_{\infty,t} 
\left\|F\right\|_{\infty}
\left\|u\right\|_{\infty,t} 
+
\left\|l_t\right\|_{\sigma_0 \infty,t}
c_{\sigma,\sigma_0}(G,t)  \nonumber
\\
&\,\left\|x\right\|_{\sigma \infty,t-1}, \left\|s_t\right\|_{\infty,t} 
\left\|F\right\|_{\infty}
\left\|u\right\|_{\infty,t} \nonumber
+ \left\|g_t\right\|_{\sigma \infty,t} \nonumber
\\
&\,\left\|x\right\|_{\sigma \infty,t-1}\} \big \} \nonumber
\\
\leq
&\left\|s_t\right\|_{\infty,t} 
\left\|F\right\|_{\infty}
\left\|u\right\|_{\infty,t} 
+
\psi(t) \left\|x\right\|_{\sigma \infty,t-1}, \label{Thm2P2}
\end{align}
where $\psi(t)$ is defined in $(\ref{psi})$, and $(\ref{Thm2P2})$ is by $(\ref{psi})$. Next, by applying the Gr\"{o}nwall-Bellman Lemma \cite{Qin_book_16} on $(\ref{Thm2P2})$, it is true that
\begin{align}
& \|x\|_{\sigma \infty,t_i} \nonumber
\\
\leq & 
\left(\prod_{t=t_{i-1}+1}^{t_i} \psi(t)\right) \|x\|_{\sigma \infty,t_{i-1}}
+ \left\|s_{t_i}\right\|_{\infty,t_i} \|F\|_{\infty} \|u\|_{\infty,t_i} \nonumber
\\
&\quad +
\sum_{t=t_{i-1}+1}^{t_i-1}\left[\prod_{j=t+1}^{t_i} \psi(j) \left\|s_t\right\|_{\infty,t} \|F\|_{\infty} \|u\|_{\infty,t}\right] \nonumber
%-----------------
\\
\leq & 
\rho^{t_i-t_{i-1}} \|x\|_{\sigma \infty,t_{i-1}}
+ \left\|s_{t_i}\right\|_{\infty,t_i} \|F\|_{\infty} \|u\|_{\infty,t_i} \nonumber
\\
&\quad +
\sum_{t=t_{i-1}+1}^{t_i-1}\left[\rho^{t_i-t} \left\|s_t\right\|_{\infty,t} \|F\|_{\infty} \|u\|_{\infty,t_i}\right] \label{Thm2P3}
\\
\leq & 
\rho^{t_i-t_{i-1}} \|x\|_{\sigma \infty,t_{i-1}}
+ \left\|s_{t_i}\right\|_{\infty,t_i} \|F\|_{\infty} \|u\|_{\infty,t_i} \nonumber
\\
&\quad +
\sum_{t=t_{i-1}+1}^{t_i-1}\left[\max_{t\in\left[t_{i-1}+1,t_i\right]}\left(  \rho^{t_i-t}  \left\|s_{t}\right\|_{\infty,t}\right) \|F\|_{\infty} \|u\|_{\infty,t_i}\right] \nonumber
%-----------------
\\
\leq & 
\rho^{t_i-t_{i-1}} \|x\|_{\sigma \infty,t_{i-1}}  \nonumber
\\
&+
\left(t_i-t_{i-1}\right)\left[\max_{t\in\left[t_{i-1}+1,t_i\right]}\left(  \rho^{t_i-t}  \left\|s_{t}\right\|_{\infty,t}\right) \|F\|_{\infty} \|u\|_{\infty,t_i}\right] \nonumber 
%-----------------
\\
\leq & 
\rho^{t_i-t_{i-1}} \|x\|_{\sigma \infty,t_{i-1}}+ \beta \|u\|_{\infty,t_i} \label{Thm2P4},
\end{align}
where $(\ref{Thm2P3})$ is by $(\ref{Thm2C1})$ and by $\|u\|_{\infty,t} \leq \|u\|_{\infty,t_i}\,\forall t \leq t_i$, and $(\ref{Thm2P4})$ is by $(\ref{bar_t})$ and $(\ref{beta})$. By considering $\rho \in \left(\sigma^{-1},1\right)$ and applying the Gr\"{o}nwall-Bellman Lemma \cite{Qin_book_16} on $(\ref{Thm2P4})$, $\forall a < j \leq i$ it is true that
\begin{align}
\|x\|_{\sigma \infty,t_j} \leq & \rho^{t_j-t_a}\|x\|_{\sigma \infty,t_a} + \sum_{k=a+1}^j \rho^{t_j-t_k} \beta \|u\|_{\infty,t_k}, \nonumber
\\
\leq & \rho^{t_j-t_a}\|x\|_{\sigma \infty,t_a} + \frac{\beta}{1-\rho}  \|u\|_{\infty,t_k}, \label{Thm2P5}
\\
\leq & \rho^{t_j-t_a}\|x\|_{\sigma \infty,t_a} + \frac{\beta}{1-\rho}  \|u\|_{\infty,t_i}, \label{Thm2P6}
\end{align}
where $(\ref{Thm2P5})$ is by $\rho^{t_j-t_k}\leq \rho$ and $(\ref{Thm2P6})$ is by $\|u\|_{\infty,t_k} \leq \|u\|_{\infty,t_i}$ respectively. Next, by choosing $a$ such that $\|x\|_{\sigma \infty,t_a}=0$, the following inequality holds:
\begin{align}
\|x\|_{\sigma \infty,t_j} \leq \frac{\beta}{1-\rho}\|u\|_{\infty,t_i}. \label{Thm2Pa}
\end{align}
Therefore for all $t \in \left[t_{j-1}+1,t_j\right]$, 
\begin{align}
\sigma^{-\left(t_j-t\right)}\left|x(t)\right| \leq & \frac{\beta}{1-\rho}\|u\|_{\infty,t_i}, \label{Thm2Pb}
\\
\therefore \quad \left|x(t)\right| \leq & \sigma^{\left(t_j-t\right)} \frac{\beta}{1-\rho}\|u\|_{\infty,t_i}, \nonumber
\\
\therefore \quad \|x\|_{\infty,\left[t_{j-1}+1,t_j\right]} \leq & \sigma^{\bar{t}-1}\frac{\beta}{1-\rho}\|u\|_{\infty,t_i}, \label{Thm2P7}
\\
\therefore \quad \|x\|_{\infty,\left[t_{j-1}+1,t_j\right]} \leq & c\|u\|_{\infty,t_i}, \label{Thm2P8}
\\
\therefore \quad \|x\|_{\infty,t_i} \leq & c\|u\|_{\infty,t_i}, \label{Thm2P9}
\end{align}
where $(\ref{Thm2Pb})$ is by $(\ref{Thm2Pa})$ and Definition \ref{sigmapsignalnorm}, $(\ref{Thm2P7})$ is by $(\ref{bar_t})$ and Definition \ref{sigmapsignalnorm}, $(\ref{Thm2P8})$ is by $(\ref{c})$, and $(\ref{Thm2P9})$ is by $\|x\|_{\infty,t_i} \leq \sup_{j \leq i}\|x\|_{\infty,\left[t_{j-1}+1,t_j\right]}$ respectively. Hence, the claim follows by noting $x = \Sigma u$. \hfill $\blacksquare$
%------------------------------------------------- Theorem 4.1 proof ------------------------------------------------------------------------------------

\begin{center} \textit\bfseries{Appendix D \\ Proof of Corollary \ref{notweakstab}} \end{center}

\par  Let $t \in \left[t_{i-1}+1,t_i\right]$ for some $i \in \mathbb{Z}$. By Fig. \ref{feedbacksys}, we have
\begin{align}
x(t) &= f_tu+g_t \mathcal{T}x, \nonumber
\\
|x(t)|& 
\leq 
\left\|F\right\|_{\infty}
\left\|u\right\|_{\infty,t} 
+
\left\|g_t\right\|_{\sigma \infty,t}
\left\|\mathcal{T}x\right\|_{\sigma \infty,t} \text{(by Lemma \ref{snapshot_norm_bound})}\nonumber
\\
|x(t)|
&\leq
\left\|F\right\|_{\infty}
\left\|u\right\|_{\infty,t} 
+
\left\|g_t\right\|_{\sigma \infty,t}
\left\|x\right\|_{\sigma \infty,t-1}. \label{Thm3P2}
\end{align}
Next, by the property of $\ell_{\sigma\infty}$-semi norm $\left\|x\right\|_{\sigma \infty,t} \leq 
\max\left\{\sigma^{-1}\left\|x\right\|_{\sigma \infty,t-1},\left|x(t)\right|\right\}$, and by $(\ref{Thm2P1})$ and $(\ref{Thm3P2})$, we have
\begin{align}
&\left\|x\right\|_{\sigma \infty,t}\leq
\max\{\sigma^{-1} \left\|x\right\|_{\sigma \infty,t-1}, \nonumber
\\
&\quad \left\|s_t\right\|_{\infty,t} 
\left\|F\right\|_{\infty}
\left\|u\right\|_{\infty,t} 
+
\left\|l_t\right\|_{\sigma_0 \infty,t}
c_{\sigma,\sigma_0}(G,t) \left\|x\right\|_{\sigma \infty,t-1}\} \label{Thm3P3}
\\
& \hspace{0cm}\text{and} \nonumber
\\
&\left\|x\right\|_{\sigma \infty,t}\leq
\max\{\sigma^{-1} \left\|x\right\|_{\sigma \infty,t-1}, \nonumber
\\
&\quad 
\left\|F\right\|_{\infty}
\left\|u\right\|_{\infty,t} 
+
\left\|g_t\right\|_{\sigma \infty,t}\left\|x\right\|_{\sigma \infty,t-1}\}. \label{Thm3P4}
\end{align}
By $(\ref{globalpsi})$, $(\ref{globalgamma})$, $(\ref{Thm3P3})$ and $(\ref{Thm3P4})$,  we have
\begin{align}
&\hspace{-2.5cm}\left\|x\right\|_{\sigma \infty,t}\leq \gamma(t)
\left\|u\right\|_{\infty,t} 
+
\hat{\psi}(t)\left\|x\right\|_{\sigma \infty,t-1}. \label{Thm3P5}
\end{align}
Next, by applying the Gr\"{o}nwall-Bellman Lemma \cite{Qin_book_16} on $(\ref{Thm3P5})$, it is true that for $t \in \left[t_{i-1}+1,t_i\right]$, we have
\begin{align}
& \|x\|_{\sigma \infty,t} \nonumber
\\
\leq & 
\left(\prod_{\tau=t_{i-1}+1}^{t} \hat{\psi}(\tau)\right) \|x\|_{\sigma \infty,t_{i-1}}
+ \gamma(t) \|u\|_{\infty,t} \nonumber
\\
&\quad +
\sum_{\tau=t_{i-1}+1}^{t-1}\left[\prod_{j=\tau+1}^{t} \hat{\psi}(j) \gamma(\tau) \|u\|_{\infty,\tau}\right] \nonumber
%-----------------
\\
\leq & 
\left(\sigma\rho\right)^{t_i-t}\rho^{t-t_{i-1}} \|x\|_{\sigma \infty,t_{i-1}}
+ \gamma(t) \|u\|_{\infty,t} \nonumber
\\
&\quad +
\sum_{\tau=t_{i-1}+1}^{t-1}\left[\left(\sigma\rho\right)^{t_i-t}\rho^{t-\tau} \gamma(\tau) \|u\|_{\infty,t}\right] \label{Thm3P6.1}
\\
\leq & 
\begin{cases}
\rho^{t_i-t_{i-1}} \|x\|_{\sigma \infty,t_{i-1}}
+ \hat{\beta} \|u\|_{\infty,t_i}, & \text{if } t = t_i,
\\
\left( \sigma \rho \right)^{\bar{t}} \|x\|_{\sigma \infty,t_{i-1}}
+ \hat{\beta} \|u\|_{\infty,t}, & \text{if } t \in \left[t_{i-1}+1,t_i\right],
\end{cases} \label{Thm3P6}
\end{align}
where $(\ref{Thm3P6.1})$ is by $\hat{\psi}(t)\geq \sigma^{-1}$ for all $t \in \mathbb{Z}$ and $(\ref{Thm3C1})$, $(\ref{Thm3P6})$ is by $(\ref{Thm3C1})$ and $(\ref{globalbeta})$. By applying the Gr\"{o}nwall-Bellman Lemma \cite{Qin_book_16} on $(\ref{Thm3P6})$, $\forall a < j \leq i-1$ it is true that
\begin{align}
\|x\|_{\sigma \infty,t_j} \leq & \rho^{t_j-t_a}\|x\|_{\sigma \infty,t_a} + \sum_{k=a+1}^j \rho^{t_j-t_k} \hat{\beta} \|u\|_{\infty,t_k}, \nonumber
\\
\leq & \rho^{t_j-t_a}\|x\|_{\sigma \infty,t_a} + \frac{\hat{\beta}}{1-\rho}  \|u\|_{\infty,t_k}, \label{Thm3P7}
\\
\leq & \rho^{t_j-t_a}\|x\|_{\sigma \infty,t_a} + \frac{\hat{\beta}}{1-\rho}  \|u\|_{\infty,t_j}, \label{Thm3P8}
\end{align}
where $(\ref{Thm3P7})$ is by $\rho^{t_j-t_k}\leq \rho$ and $(\ref{Thm3P8})$ is by $\|u\|_{\infty,t_k} \leq \|u\|_{\infty,t_j}$ respectively. Next, by choosing $a$ such that $\|x\|_{\sigma \infty,t_a}=0$, the following inequality holds:
\begin{align}
\|x\|_{\sigma \infty,t_{i-1}} \leq \frac{\hat{\beta}}{1-\rho}\|u\|_{\infty,t_{i-1}}. \label{Thm3P9}
\end{align}
Therefore by $(\ref{Thm3P6})$ and $(\ref{Thm3P9})$, for all $t \in \left[t_{i-1}+1,t_i\right]$, 
\begin{align}
\|x\|_{\sigma \infty,t} \leq & \left( \sigma \rho \right)^{\bar{t}} \left\|x\right\|_{\sigma \infty,t_{i-1}} + \hat{\beta}\|u\|_{\infty,t}, \nonumber
\\
\leq & \left( \sigma \rho \right)^{\bar{t}} \frac{\hat{\beta}}{1-\rho}\left\|u\right\|_{\infty,t_{i-1}}+\hat{\beta} \left\|u\right\|_{\infty,t}, \nonumber
\\
\leq & \left(\frac{\left( \sigma \rho \right)^{\bar{t}}}{1-\rho}+1\right)\hat{\beta} \left\|u\right\|_{\infty,t}, \nonumber
\\
\left|x(t)\right| \leq & \hat{c} \left\|u\right\|_{\infty,t}, \nonumber
\\
\left\|x\right\|_{\infty,\left[t_{i-1}+1,t_i\right]} \leq & \hat{c} \left\|u\right\|_{\infty,t}. \label{Thm3P10}
\end{align}
By $(\ref{Thm3P10})$ and $\cup_{i \in \mathbb{Z}}\left[t_{i-1}+1,t_i\right] = \mathbb{Z}$, we have $\left\|x\right\|_{\infty,t} \leq \hat{c} \left\|u\right\|_{\infty,t}$ and thus $\left\|\Sigma\right\|_{\infty,t} \leq \hat{c}$ for all $t \in \mathbb{Z}$. \hfill $\blacksquare$
%------------------------------------------------- Lemma 4.4 proof ------------------------------------------------------------------------------------
\begin{center} \textit\bfseries{Appendix E \\ Proof of Lemma \ref{less_conservative}} \end{center}
\par Since the sufficient condition $(\ref{CS_specical_c_N_c1})$ is a special case of the sufficient conditions $(\ref{Thm2C1})$ and $(\ref{Thm3C1})$ with $\sup_{t\in\mathbb{Z}}\|s_t\|_{\infty} < \infty$, $\sup_{t\in\mathbb{Z}}\|l_t\|_{\sigma_0 \infty} < \infty$,  $\left\{t_i\right\}=\mathbb{Z}$, and the $N$-width average variation rate of $G$ is bounded, it is true that $(\ref{CS_specical_c_N_c1}) \Rightarrow (\ref{Thm2C1})$ and $(\ref{CS_specical_c_N_c1}) \Rightarrow (\ref{Thm3C1})$. Therefore, to prove that $(\ref{Thm2C1})$ and $(\ref{Thm3C1})$ hold whenever $(\ref{Zames_cons_c1})$ holds and there exist cases where $(\ref{Thm2C1})$ and $(\ref{Thm3C1})$ hold while $(\ref{Zames_cons_c1})$ does not hold, it suffices to prove (i) $(\ref{Zames_cons_c1}) \Rightarrow (\ref{CS_specical_c_N_c1})$, and (ii) $(\ref{CS_specical_c_N_c1}) \nRightarrow (\ref{Zames_cons_c1})$. 
\par (i) Let $N=1$, and thus $\bar{\bar{d}}_{\sigma,N}=\bar{\bar{d}}_{\sigma,1}$ by $(\ref{bar_bar_d})$ and $(\ref{bar_bar_d_1})$. Since $\bar{d}_{\sigma,N}(G) \triangleq \sup_{t \in \mathbb{Z}}d_{\sigma,N}(G)(t)$, we have $(\ref{Zames_cons_c1}) \Leftrightarrow (\ref{CS_specical_c_N_c1})$ by $\bar{\bar{d}}_{\sigma,N}(G)=\bar{\bar{d}}_{\sigma,1}(G)$ and Definitions \ref{snapshot} and \ref{Nwidth}. Therefore it is true that $(\ref{Zames_cons_c1}) \Rightarrow (\ref{CS_specical_c_N_c1})$.
\par (ii) Consider a case where $N>1$, $N\left(\frac{\sigma_0}{\sigma}\right)^{1-N}>1$, and a time $\tau \in \mathbb{Z}$ such that 
\begin{align*}
\left\|\nabla g_{\tau}\right\|_{\sigma \infty} \in \left(\bar{\bar{d}}_{\sigma,1}(G),N\bar{\bar{d}}_{\sigma,N}(G)\right]
\end{align*}
and
\begin{align*}
\sum_{i = t-N+1}^t \left\|\nabla g_{i}\right\|_{\sigma \infty}\leq N\bar{\bar{d}}_{\sigma,N},\forall t \in \mathbb{Z}.
\end{align*} 
Since $N\left(\frac{\sigma_0}{\sigma}\right)^{1-N}>1$, it is true that $N\bar{\bar{d}}_{\sigma,N} > \bar{\bar{d}}_{\sigma,1}$ and $\left(\bar{\bar{d}}_{\sigma,1},N\bar{\bar{d}}_{\sigma,N}\right]$ is not empty. Next, by $\left\|\nabla g_{\tau}\right\|_{\sigma \infty} > \bar{\bar{d}}_{\sigma,1}$ $(\ref{Zames_cons_c1})$ does not hold. On the other hand, by $\bar{d}_{\sigma,N}(G)\triangleq\sup_{t \in \mathbb{Z}}d_{\sigma,N}(G)(t)$ and by $\sum_{i = t-N+1}^t \left\|\nabla g_{i}\right\|_{\sigma \infty}\leq N\bar{\bar{d}}_{\sigma,N}(G)$, inequality $(\ref{CS_specical_c_N_c1})$ holds. Then it is true that $(\ref{CS_specical_c_N_c1}) \nRightarrow (\ref{Zames_cons_c1})$.
\par By (i), (ii), $(\ref{CS_specical_c_N_c1}) \Rightarrow (\ref{Thm2C1})$ and $(\ref{CS_specical_c_N_c1}) \Rightarrow (\ref{Thm3C1})$, it is true that $(\ref{Zames_cons_c1}) \Rightarrow (\ref{Thm2C1})$, $(\ref{Thm2C1}) \nRightarrow (\ref{Zames_cons_c1})$, $(\ref{Zames_cons_c1}) \Rightarrow (\ref{Thm3C1})$, and $(\ref{Thm3C1}) \nRightarrow (\ref{Zames_cons_c1})$. Therefore the lemma is proved. \hfill $\blacksquare$
%------------------------------------------------- References ---------------------------------------------------------------------------------
\bibliographystyle{IEEEtran}
\bibliography{IEEEabrv,CoDIT19}

%------------------------------------------------- Biographies ---------------------------------------------------------------------------------
\vskip -2\baselineskip plus -1fil
\begin{IEEEbiography}[{\includegraphics[width=1in,height=1.25in,clip,keepaspectratio]{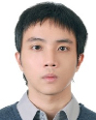}}]{Yu-Chen Sung}
	
	(S'16) was born in Kaohsiung, Taiwan, on February 16, 1984. He received the B.S. degree in electrical engineering from National Taiwan University, Taipei, Taiwan, in 2009, the M.S. and Ph.D. degrees in electrical engineering from the University of Southern California, Los Angeles, CA, USA, in 2011 and 2019, respectively.
	
	\par He is currently a Post-doctorate at the Ming Hsieh Department of Electrical Engineering of the University of Southern California, Los Angeles, CA, USA. His research interests include $H_{\infty}$ loop-shaping, data-driven control, robust control, and adaptive control.
	
\end{IEEEbiography}
\vskip -2\baselineskip plus -1fil
\begin{IEEEbiography}[{\includegraphics[width=1in,height=1.25in,clip,keepaspectratio]{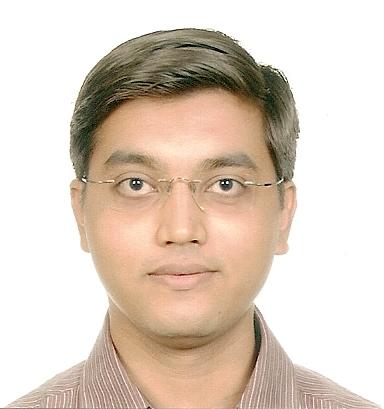}}]{Sagar V. Patil}
	
	was born in Pune, MH, India, on February 10, 1986. He received the B.Tech. degree in electrical engineering from the College of Engineering Pune, MH, India, in 2008, the M.S. and Ph.D. degrees in electrical engineering from the University of Southern California, Los Angeles, CA, USA, in 2011 and 2016, respectively.
	
	\par From 2008 to 2009, he was a Graduate Engineering Trainee at the Wipro Consumer Care \& Lighting, Pune, MH, India. From 2016 to 2017, he was a Post-doctorate at the Ming Hsieh Department of Electrical Engineering of the University of Southern California, Los Angeles, CA, USA. From 2017 to 2018, he was a Research Associate at the Energy \& Environment Directorate of the Pacific Northwest National Laboratory, Richland, WA, USA. Since 2018 he has been with the Bajaj Auto, Pune, MH, India where he is presently a Control Systems R\&D Engineer. His research interests include adaptive, nonlinear, robust, \& data-driven control, $H_{\infty}$ loop-shaping, bumpless controller switching algorithms, powertrain (HEV) \& aftertreatment modeling, stochastic distribution control, traffic flow modeling \& control, connected \& automated vehicles, and estimation.
	
\end{IEEEbiography}
\vskip -2\baselineskip plus -1fil
\begin{IEEEbiography}[{\includegraphics[width=1in,height=1.25in,clip,keepaspectratio]{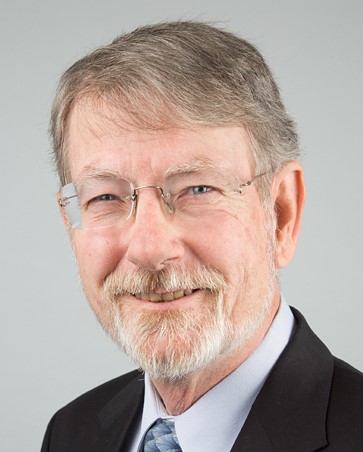}}]{Michael G. Safonov}
	
	(M'73--S'76--M'77--SM'82--F'89-- LF'14) was born in Pasadena, California. He received the B.S., M.S., Engineer, and Ph.D. degrees in electrical engineering from the Massachusetts Institute of Technology, Cambridge, MA in 1971, 1971, 1976 and 1977, respectively. 
	
	\par From 1972 to 1975 he served with the U.S. Navy as Electronics Division Officer aboard the aircraft carrier USS Franklin D. Roosevelt (CVA-42). Since 1977 he has been with the University of Southern California where he is presently Professor Emeritus of Electrical Engineering. He has been Consultant to The Analytic Sciences Corp., Honeywell Systems and Research Center, Systems Control, Systems Control Technology, Scientific Systems, United Technologies, TRW, Northrop Aircraft, Hughes Aircraft and others.   His consulting and university research activities have involved him flight control system design studies in which modern robust multivariable control techniques were applied to a variety of aircraft including the CH-47 Chinook helicopter (Analytic Sciences Corp., 1976), the NASA HiMAT aircraft (Honeywell/USC, 1980) and the F/A-18 Hornet (Northrop, 1987-1991). During the academic year 1983-1984 he was a Senior Visiting Fellow with the Department of Engineering, Cambridge University, England, and in summer 1987 he held a similar appointment at Imperial College of Science and Technology, London, England and in 1990-1991 at Caltech, Pasadena, CA. He has authored or co-authored more than three hundred journal and conference papers, and the book \emph{Stability and Robustness of Multivariable Feedback Systems} (MIT Press, 1980)  and  \emph{Safe Adaptive Control: Data-driven Stability Analysis and Robust Synthesis} (Springer-Verlag, 2011).  Additionally, he is co-author of the MATLAB Robust Control Toolbox (Natick, MA: MathWorks). His research interests include robust control, adaptive control and nonlinear system theory, with applications to aerospace control design problems. 
	
	\par Prof. Safonov has served as an Associate Editor of \emph{IEEE Trans. on Automatic Control} and \emph{Systems and Control Letters} and is presently on the editorial board of \emph{International Journal of Robust and Nonlinear Control}. From 1993 to 1995, he was Chair of the Awards Committee of the American Automatic Control Council. He is an IFAC Fellow.
	
\end{IEEEbiography}

%---------------------------------------------------------------------------------------------------------------------------------------------------

\end{document}